\documentclass[pdflatex,sn-mathphys-num]{sn-jnl}
\usepackage{subfig}%
\usepackage{graphicx}%
\usepackage{multirow}%
\usepackage{amsmath,amssymb,amsfonts}%
\usepackage{amsthm}%
\usepackage{mathrsfs}%
\usepackage[title]{appendix}%
\usepackage{xcolor}%
\usepackage{textcomp}%
\usepackage{manyfoot}%
\usepackage{booktabs}%
\usepackage{algorithm}%
\usepackage{algorithmicx}%
\usepackage{algpseudocode}%
\usepackage{listings}%

\usepackage{graphicx}
\usepackage{newtxtext}
\usepackage{newtxmath}
\usepackage{natbib}
\usepackage{hyperref}
\usepackage{mathtools}
\usepackage{bm}
\usepackage{comment}
\hypersetup{
    colorlinks = true,
    urlcolor   = blue,
    citecolor  = black,
}

\theoremstyle{thmstyleone}%
\theoremstyle{thmstyletwo}%

\theoremstyle{thmstylethree}%

\newcommand{\bc}{{ \bm c}}

\newcommand{\bx}{{ \bm{x}}}

\newcommand{\bbf}{{ \bm{f}}}

\newcommand{\beps}{{ \bm{\epsilon}}}
\newcommand{\bg}{{ \bm g}}

\newcommand{\bu}{{ \bm{u}}}

\newcommand{\defeq}{\vcentcolon=}

\newcommand{\mat}[1]{{\mathbf{#1}}}

\newcommand{\Rey}{Re}

\newcommand{\mA}{\mat{A}}

\newcommand{\mI}{\mat{I}}
\newcommand{\mL}{\mat{L}}
\newcommand{\mM}{\mat{M}}

\newcommand{\mR}{\mat{R}}
\newcommand{\mQ}{\mat{Q}}

\newcommand{\mU}{\mat{U}}

\newcommand{\Laplace}{{\mathbf{\Delta}}}
\newcommand{\revone}[1]{{{#1}}}

\begin{document}
\title[Operator theoretic causality analysis]{
\revone{Operator theoretic causality analysis of fluid flows using linearized dynamics}
}

\author[1]{\fnm{Ankit} \sur{Srivastava}}\email{asriva13@illinoistech.edu}

\author[1]{\fnm{V\'{i}ctor} \sur{Jim\'{e}nez-Fern\'{a}ndez}}\email{vjimenezfernandez@hawk.illinoistech.edu}

\author[1]{\fnm{Louis N.} \sur{Cattafesta III}}\email{lcattafestaiii@illinoistech.edu}

\author*[1]{\fnm{Scott T. M.} \sur{Dawson}}\email{sdawson5@illinoistech.edu}

\affil*[1]{\orgdiv{Mechanical, Materials, and Aerospace Engineering Department}, \orgname{Illinois Institute of Technology}, \orgaddress{\street{10 W. 32$^{\text{nd}}$ St.}, \city{Chicago}, \postcode{60616}, \state{IL}, \country{USA}}}


\abstract{
This paper presents an operator-theoretic framework, Linear Operator Causality Analysis (LOCA), for analyzing causality in linearized dynamical systems, focusing here on fluid flows. \revone{Our proposed approach, which can be characterized as a special case of Dynamic Causal Effect (DCE) analysis, utilizes} the matrix exponential of linearized differential equations 
 \revone{to determine} causal relationships between system modes at any future time. We further develop an upper bound that quantifies the presence and extent of global causality across all time horizons. This approach provides a physics-based alternative to \revone{data-driven} statistical and information-theoretic causality measures such as Granger causality and transfer entropy. Unlike these data-driven techniques that infer causality from time-series data,  LOCA leverages the linearized governing equations, yielding a \revone{physically-motivated } and interpretable measure of causal interactions. We \revone{identify the conditions under which} LOCA gives equivalent results to data-driven causality analysis methods, and further discuss connections to key system properties such as controllability, observability, and graph-theoretic transitive closure. To complement this operator-based approach, we introduce a data-driven methodology akin to Dynamic Mode Decomposition (DMD) that estimates causal connections directly from time series data by approximating the matrix exponential. We argue that LOCA also mitigates common issues in data-driven causality analyses, such as misleading inferences due to correlated variables or state truncation. We demonstrate our method on two fluid flow examples: linearized Couette flow, and a nonlinear wake flow featuring chaotic dynamics. In both cases, we demonstrate how our framework captures both direct and indirect causal interactions among flow structures.
}


\maketitle

\section{Introduction}\label{sec:Intro}

Fluid flows often exhibit complex interactions across a wide range of spatial and temporal scales. The intricate coupling between different modes and scales leads to instabilities, flow separation, and turbulence, which are challenging to predict and control. Understanding the causal relationships governing these interactions is essential for advancing fluid dynamics research and its applications.

The concept of causality has been a subject of intense scrutiny across multiple disciplines in science and engineering. The counterfactual approach, developed in philosophy \citep{lewis1973causation} and statistics \citep{rubin1974estimating}, posits that an event A causes event B if B would not have occurred had A not happened. In contrast, probabilistic frameworks in epidemiology \citep{suppes1973probabilistic} and computer science \citep{pearl2014probabilistic,pearl2000models} define causality in terms of probability distributions, where A causes B if the probability of B given A is greater than the probability of B alone. The field of economics has contributed the notion of Granger causality \citep{granger1969investigating}, which asserts that A causes B if past values of A improve predictions of B beyond what past values of B alone can predict, a concept particularly useful for time-series data. Social sciences have employed structural equation modeling \citep{bollen2014structural}, a statistical technique that combines factor analysis and multiple regression to test complex causal hypotheses. Epidemiology has utilized the Bradford Hill criteria \citep{hill1965environment} to evaluate causal evidence in observational studies. At the same time, biology and social sciences have emphasized mechanism-based causality \citep{machamer2000thinking}, focusing on identifying and understanding the underlying processes that connect cause and effect. This interest has also led to the development of a range of causality quantification metrics. Some examples are ATE (Average Treatment Effect) \citep{imbens2015causal}, Granger causality $F$-test \citep{granger1969investigating} and the related PDC (Partial Directed Coherence) \citep{baccala2001partial}, the Causal Impact method \citep{brodersen2015inferring}, estimators based upon the Rubin Causal Model \citep{imbens2015causal}, metrics coming from Information theory such as Directed Information (DI) \citep{massey1990causality} and Transfer Entropy (TE) \citep{schreiber2000measuring}, and those coming from Ecology and climate science such as Cross Convergent Mapping (CCM) \citep{sugihara2012detecting} to detect causality in complex, nonlinear systems. In addition, there are dedicated methods for causality analysis in deterministic systems in physics and signal processing, which employ the Hilbert transform and the associated Kramers-Kronig relations \citep{Kronig1926, kramers1927diffusion,srivastava2020causality}. 

\revone{
A complementary dynamical-systems perspective is provided by the framework of dynamical causal effects (DCEs)~\citep{Smirnov2014,Smirnov2022}. In this framework, causal influence is characterized through the response of one part of a dynamical system to a prescribed variation in another, with the response evaluated after propagation over a specified temporal horizon. The framework encompasses deterministic and stochastic systems and allows different causal quantifiers to be constructed through different choices of initial variation and response measure. This perspective is particularly relevant when the governing dynamics, rather than only observational time series, are available. The present work builds on this general dynamical-systems viewpoint by focusing on operator-based causal measures for linear and linearized fluid systems, together with their computation from known operators or operator estimates obtained from data.
}

Identifying and quantifying causal relationships in fluid-dynamic systems represents an important frontier in understanding complex flow phenomena. Early attempts to identify causality in fluid systems relied primarily on statistical and information-theoretic methods. Whereas statistical and information-theoretic methods infer causality from observed data, our framework quantifies causal influence directly from the linearized governing equations. This operator-theoretic formulation, a physics-based alternative, not only complements or guides data-driven analyses, but also opens up new avenues for understanding complex fluid systems.
The pioneering work of \citet{favre1957space} introduced space-time correlations to fluid dynamics, providing a framework for identifying potential cause-effect relationships through time delays between signals. Cross-correlation analysis identifies time delays between signals to suggest potential cause-and-effect relationships. While conceptually simple and computationally efficient, this approach only captures linear relationships and cannot distinguish between direct causal links and common-cause relationships. \citet{bendat2011random} applied the coherence function, which measures the frequency-dependent correlation between signals, to fluid dynamics problems. This technique proved valuable for analyzing aeroacoustic phenomena and identifying frequency bands where different flow variables exhibit potential causal relationships.

The past fifteen years have witnessed advancements in applying causality concepts to fluid dynamics, with methods increasingly borrowed from statistics, information theory, and dynamical systems theory. \citet{Tissot2014GrangerCI} used Granger causality to study the dynamics and the energy redistribution between scales and components in wall-bounded turbulent flows. \citet{san2016preliminary} demonstrated, through rigorous causality analysis, that the streak-roll regeneration cycle in turbulent channel flows comprises two mutually causal subcycles interconnected by wall-normal and spanwise velocity interactions. \citet{lozano2020causality} showed that energy-containing eddies in wall turbulence exhibit self-similar causal interactions across buffer and logarithmic layers regardless of eddy size. \citet{Bae2019CausalAO} used transfer entropy to show that the self-sustaining process in the logarithmic layer of wall-bounded turbulence is predominantly unidirectional rather than cyclic, with log-layer motions primarily maintained by energy extraction from the mean shear that governs the system's dynamics and time-scales. \citet{LozanoDurn2020CauseandeffectOL} demonstrated through cause-and-effect analysis based on interventions in direct numerical simulations that transient growth, particularly the Orr/push-over mechanism induced by spanwise variations of the base flow, is sufficient for sustaining realistic wall turbulence even when exponential instabilities, neutral modes, and parametric instabilities of the mean flow are suppressed. \citet{lozano2022information} proposed an information-theoretic approach, leveraging Shannon entropy, to quantify causality with applications to turbulence. More generally, \citet{cerbus2013information} also considered quantifying information content of a turbulent fluid flow using Shannon entropy. 
\citet{martinez2023causality} applied conditional transfer entropy to identify causal relationships among reduced-order model modes, revealing vortex-breaker modes as the primary drivers of large-scale structures in flow around a wall-mounted square cylinder. \citet{Massaro2023OnTP} demonstrated the effectiveness of transfer entropy in capturing directional causality in turbulent dynamical systems, highlighting its sensitivity to Markovian order. \citet{lopez2024linear} applied linear and nonlinear Granger causality analyses to turbulent duct flows, revealing that linear interactions dominate causal interactions between modes sharing the same symmetry properties, while nonlinear effects govern causal interactions between symmetric and antisymmetric coherent structures. \citet{MartnezSnchez2024DecomposingCI} introduced SURD (Synergistic-Unique-Redundant Decomposition), a method that decomposes causality into synergistic, unique, and redundant components, providing a more robust measure of causal inference in complex nonlinear systems.

These prior works \revone{studying causal mechanisms within fluid flows generally} focus on assessing causality using data and/or intrusive manipulation of simulations. Conversely, there has been minimal prior work attempting to analyze causality of fluid flows directly from the known governing equations. More generally, such equation based analyses is possible for Granger causality of linear systems \citep{barnett2015granger}, for information transfer in two-state stochastic nonlinear systems \citep{liang2005information,liang_information_2016} ( discussed in the context of turbulent flows in \citet{san2016preliminary}) \revone{and more generally using the DCE framework discussed previously \cite{Smirnov2014,Smirnov2022}}.

\revone{A drawback of data-driven causality analysis methods is that achieving statistical convergence often requires consideration of only a limited number of measurements, states or modes, which may obscure some}
causal pathways inherent in high-dimensional fluid systems. To address this gap, we propose using properties of the underlying governing equations and linearizations to deduce causal interactions from a physics-based, rather than data-driven, perspective.
Note that in other contexts, linear operator-based methods help capture structures and dynamics of even highly turbulent flows \citep{schoppa2002coherent,del2006linear,mckeon2010critical,abreu2020spectral}. 
Data-driven causality analyses typically seek causal relationships between predefined measurements or spatial functions. By contrast, we will show how operator-driven analysis, \revone{which turns out to correspond to a specific form of DCE,} can enable the identification of optimally causal pathways from full state information. 

The remainder of this paper is organized as follows. In Sect.~\ref{sec:causality}, we formulate three definitions and/or quantifications of causality under our LOCA framework: immediate causality, delayed causality, and a bound on global causality. Sect.~\ref{sec:connections} \revone{describes the connection between our causality framework and DCE and other specific } data-driven methods such as Granger causality and Transfer Entropy, and further  explores its relationships with graph theory and controllability and observability of dynamical systems. Section~\ref{sec:causal-inference} develops a data-driven approach for inferring causal relationships directly from time-series data, directly analogous to our operator-based methodology. Section~\ref{sec:resultsCouette} demonstrates the application of our methodology to linearized planar (Couette) flow, while Sect.~\ref{sec:results-wake} considers it's application to a chaotic wake flow. Finally, Sect.~\ref{sec:conclusions} summarizes our findings and discusses potential directions for future research.

\section{Definitions and quantification of causality}
\label{sec:causality}
Here, we provide a \revone{specific} notion for quantifying causality, which is particularly useful for situations where the underlying governing equations are known. We first formulate this definition for a general dynamical system, before focusing on the case where the dynamics are linear or linearized. While many of the concepts discussed here are related to standard concepts in linear dynamical systems, their exposition will be instructive to contextualize our proposed methods for quantifying causality.

Suppose that the dynamical system under consideration has a state vector $\bu$, with dynamics governed by 
\begin{equation}
\label{eq:f}
    \dot \bu = \bbf(\bu),
\end{equation}
where $\bbf$ is a (nonlinear) function governing the dynamics of the system. Evolving this system from an initial condition $\bu^0 = \bu(t=0)$ over a time horizon $\tau$, we can define the finite-time propagation operator
\begin{equation}
\label{eq:g}
   \bu^\tau = \bg^\tau(\bu^0).
\end{equation}
Suppose we wish to analyze the causal mechanisms present in this system. In that case, we may analyze the effect of a specific component $u_j$ upon another state $u_i$ over a specified time horizon. Mathematically, if we denote the perturbation in state $u_j$ at a certain time $t_0$ as $\partial u_j^{0}$, then we are interested in quantifying the change $\partial u_i^{t_0+\tau}$ where $\tau>0$. In other words, we assume that the causal influence that a perturbation of state $j$ has on the value of state $i$ at a time horizon of $\tau$ time units in the future is proportional to
\begin{equation}
\label{eq:caus1}
 G^\tau_{j\to i} = \frac{\partial u_i^\tau}{\partial u_j^0} = \frac{\partial g^\tau_i(\bu^0)}{\partial u_j^0}.
\end{equation}
 This work will focus on linearized dynamics about some base state $\overline{\bu}$,  and  instead considers the linearized equivalent to Eq.~\eqref{eq:f}
 \begin{equation}
     \dot{\bu}' = \left.\frac{\partial \bbf}{\partial \bu'}\right|_{\bu = \overline{\bu}} \bu' \defeq \mA\bu',
 \end{equation}
where $\bu' = \bu -\overline{\bu}$ is the deviation of $\bu$ from the base state. Henceforth, we will omit the $\cdot '$ notation and assume that the system's state is relative to this base state.
 In this case, Eq.~\eqref{eq:g} becomes
 \begin{equation}
 \label{eq:glin}
     \bu^\tau = e^{\mA\tau}\bu^0,
 \end{equation}
and the causality definition given in Eq.~\eqref{eq:caus1} simplifies to 
\begin{equation}
\label{eq:caus3}
 G^\tau_{j\to i} = \left[  e^{\mA\tau}\right]_{ij}.
\end{equation}
While our defined notion of causality is proportional to $G^\tau_{j\to i}$, we will ultimately modify this definition to account for the statistics of the state variables, motivated by classical notions of causality addressed in Sect.~\ref{sec:granger}. In particular, if we assume that the dynamics given in Eq.~\eqref{eq:glin} also have an error term $\beps$ (which could be due to unmodeled nonlinear dynamics)
 \begin{equation}
 \label{eq:disclin}
     \bu^\tau = e^{\mA\tau}\bu^0 + \beps,
 \end{equation}
then we define a modified $F$-statistic associated with the strength of the causality from state $u_j$ to state $u_i$ by
\begin{equation}
\label{eq:F1}
    F_{j\to i} = \frac{\left[e^{\mA\tau}\right]^2_{ij}\sigma_j^2}{\sigma_{\epsilon,i}^2},
\end{equation}
where $\sigma_j^2 = \text{var}(u_j)$ and $\sigma_{\epsilon,i}^2 = \text{var}(\epsilon_i)$ are the variances of state variable $j$ and the error in the equation for variable $i$, respectively.

Note that in the absence of data, the state variance of a linear dynamical system can be computed by solving an appropriate discrete-time Lyapanov equation. In particular, for a discrete-time system of the form given by Eq.~\eqref{eq:disclin}, the covariance matrix $\mR$ (with components $R_{ij} = \text{covar}(u_i,u_j)$) is the solution to
\begin{equation}
\label{eq:dlyap}
e^{\mA\tau}
\mR
e^{\mA^*\tau} - \mR
+ \mQ = \mat{0},
\end{equation}
where $\mQ$ is the error covariance matrix (which must be given an assumed form in the absence of data). Eq.~\eqref{eq:dlyap} can be solved using standard numerical methods \citep{barraud2003numerical}, e.g.~using dlyap function within MATLAB's control system toolbox.

The state dimension is typically very large in discretized fluid flow models. Therefore, we are often interested in assessing causality not between state components directly but rather between the coefficients of a set of spatial basis functions, which only span a subspace of the original space of state variables. 
For example, if we had a set of orthonormal spatial basis functions (e.g.~POD modes) assembled as columns of a matrix $\mU$ with corresponding temporal coefficients $\bc = \mU^*\bx$, then the equivalent to Eq.~\eqref{eq:disclin} would be 
 \begin{equation}
 \label{eq:disclinr}
     \bc^\tau = e^{\tilde\mA\tau}\bc^0,
 \end{equation}
 where the equivalent linearized continuous- and discrete-time operators in POD coefficient coordinates are given by
 \begin{align}
     \tilde\mA &= \mU^*\mA \mU, \\
     e^{\tilde\mA\tau} &= \mU^*e^{\mA\tau} \mU.
 \end{align}
 See, for example, \citet{berkooz1993pod,holmes2012pod,taira2017modal} for further details regarding the properties and computation of POD. 
A reduced-order model for these dynamics can be obtained by selecting a subset of these basis functions (such as the leading $r$ POD modes), which, when assembled as columns of a matrix $\mU_r$, yield the transformations
\begin{align}
\label{eq:xtoc}
    \bc &= \mU_r^* \bx, \\
    \tilde{\mA} &= \mU_r^*\mA \mU_r.
\end{align}
 The $r\times r$ operator $\tilde{\mA}$ now provides an approximation to the dynamics of the full system, with the transformed and truncated state given by the coefficients of the first $r$ POD modes. With this in mind, we will often use $c_j$ to refer to the $j^{\mathrm{th}}$ state of the (possibly reduced) system under consideration. 
 We now discuss how several specific notions of causality can be inferred directly from the individual elements and properties of such operators.

\subsection{Immediate causality}
Our notion of causality is connected to the classical idea of counterfactual causality \citep{lewis1973causation} where causality is understood by insisting that if we deem that $c_j$ causes $c_i$, then any change in $c_j$ at some time $t=t_0$ should change the trajectory of $c_i$ for $t>t_0$. As far as immediate causality is concerned, this means that if $\dot{c}_i(t_0)\neq 0$ when ${c}_j(t_0)$ is perturbed, then the future trajectory of $c_i(t)$ will also diverge from its original trajectory. Since this local relationship is mediated by $\mathbf{A}$ through $\dot{\bc }=\mathbf{A}\bc $, immediate causality from mode $j$ to mode $i$ is ensured if $A_{ij}\neq 0$ and in this sense, immediate causality is the same as the direct coupling encoded in the dynamical matrix $\mathbf{A}$. For example, in the following system:

\begin{equation}
\mathbf{A} = \begin{bmatrix}
-1 & 1 & -1 \\
0 & -2 & 2 \\
0 & 0 & -0.02 + \mathrm{i} \\
\end{bmatrix}
\end{equation}
there are immediate causal connections in the directions $2\rightarrow 1$, $3\rightarrow 1$, and $3\rightarrow 2$ but no immediate causal connections in the reverse directions.

\subsection{Delayed causality}
Over larger time horizons, counterfactual causality is embedded in the partial derivative $G^\tau_{j\to i}$, which we call delayed causality over a time horizon $\tau$. Mathematically, if we denote the perturbation in mode $c_j$ at a particular time $t_0$ as $\partial c_j^{t_0}$, then we are interested in quantifying the change $\partial c_i^{t_0+\tau}$ where $\tau>0$. Since the systems under consideration are linear and time-invariant, we can instead talk about the quantities $\partial c_i^{\tau},\partial c_j^{0}$, and we are interested in quantifying the time delayed partial derivative $G^\tau_{j\to i}=\partial c_i^{\tau}/\partial c_j^{0}$. If $\partial c_i^{\tau}/\partial c_j^{0}=0$, then we can say that there is no causal influence of the mode $c_j$ on the state of mode $c_i$ at the future time $\tau$. As introduced above, this time-delayed partial derivative can be explicitly calculated for the linear systems under consideration and is given by the matrix exponential $\mathrm{e}^{\mathbf{A}t}$. Since $\partial c_i^{\tau}/\partial c_j^{0}=[\mathrm{e}^{\mathbf{A}\tau}]_{ij}$, the matrix $\mathrm{e}^{\mathbf{A}t}$ encapsulates the essential causal relations among the modes at any future time.

\subsection{A Global causality bound}

While the ideas discussed in the preceding sections can quantify causality instantaneously and over specified time horizon, it is often of interest to determine the presence and extent of causal influence across all time. 
In particular, since the matrix exponential is time-horizon dependent, the answer to this question cannot be provided by this matrix for a single choice of timestep. Indeed, discussions of causality based on a fixed time lag can be quite misleading. Consider, for example, the matrix $\mathbf{A}$ defined as:
\begin{equation}
\mathbf{A} = \begin{bmatrix}
0 & 0 & 0 \\
1 & 0 & 0 \\
0 & 1 & 0 \\
\end{bmatrix}.
\end{equation}
For this nilpotent matrix, we have:
\begin{equation}
\mathrm{e}^{\mathbf{A} t} = \mathbf{I} + \mathbf{A} t + \frac{(\mathbf{A} t)^2}{2!}.
\end{equation}
Substituting the powers of $\mathbf{A}$, we obtain:
\begin{equation}
[\mathrm{e}^{\mathbf{A} t}]_{31} = \frac{t^2}{2},
\end{equation}
showing a quadratic dependence on $t$. This implies that a causality metric based on $[\mathrm{e}^{\mathbf{A} t}]_{31}$, computed for a time lag $t \ll 1$, will show no causal relation from mode 1 to 3. However, a strong causal relation exists between modes 1 and 3 for $t$ on the order of 1 $\mathrm{s}$ and dominates for $t \gg 1$. 

For the following case:
\begin{equation}
\mathbf{A} = \begin{bmatrix}
0 & 0 & 0 \\
1 & 0 & 0 \\
-1 & 1 & 0 \\
\end{bmatrix},
\end{equation}
it can be shown that:
\begin{equation}
[\mathrm{e}^{\mathbf{A} t}]_{31} = \frac{t^2}{2} - t = \frac{1}{2} t (t - 2).
\end{equation}
In this situation, $\mathrm{e}^{\mathbf{A}t}$ exhibits no causality if the time lag is very small or close to 2. 
\revone{This motivates a global indicator that accounts for causal pathways over all temporal horizons. In Sect.~\ref{sec:graph}, we introduce the cancellation-free matrix $\mM_\alpha=(\alpha\mathbf{I}-|\mathbf{A}|)^{-1}$, with $\alpha>\rho(|\mathbf{A}|)$, and show that it bounds an exponentially discounted, time-integrated magnitude of the finite-time causal response. We also show that zero entries of $\mM_\alpha$ provide a certificate of causal disconnection for all time horizons.}

\revone{The same global indicator can also be used to determine whether certain modes of a dynamical system are causally disconnected from other modes for all time. In particular, as shown in Sect.~\ref{sec:graph}, a zero off-diagonal entry of $\mM_\alpha$ implies that the corresponding entry of $\mathrm{e}^{\mathbf{A}t}$ is zero for every $t\geq 0$.} 
One such scenario occurs when the matrix $\mathbf{A}$ is in block diagonal form. In this case, $\mathbf{A}^k$ is also block diagonal for each $k$, from which it follows that $\mathrm{e}^{\mathbf{A}t}$ is block diagonal for all $t$. Therefore, it follows that $|\mathbf{A}|$ is also block diagonal, and so are $\alpha\mathbf{I}-|\mathbf{A}|$ and its inverse. Thus, block diagonal $\mathbf{A}$ gives rise to causally disconnected modes over all time horizons. Note also that this logic applies if $\mathbf{A}$ and thus $|\mathbf{A}|$ are diagonal, in which case the state variables align with eigenvectors of $\mathbf{A}$. In this case, there should be no causality between different states.

\section{Connections with other methods}
\label{sec:connections}
We now discuss connections between the analysis and definitions provided in Sect.~\ref{sec:causality} and other established methods. We start by \revone{placing LOCA in the context of DCEs in Sect.~\ref{sec:loca_dce}, before discussing its relationship with other specific} notions of causality (typically applied in a data-driven manner) in Sect.~\ref{sec:granger}. We then describe connections with notions from graph theory in Sect.~\ref{sec:graph}, and observability and controllability of linear systems in Sect.~\ref{sec:bt}.

\subsection{\revone{Relationship between LOCA and dynamical causal effects}}
\label{sec:loca_dce}

\revone{
This section demonstrates that LOCA can be defined as a particular realization of the broader framework of dynamical causal effects (DCEs)~\citep{Smirnov2014,Smirnov2022}. In general, a DCE quantifies how a specified variation in one component of a dynamical system changes the subsequent behavior of another component. Its definition therefore requires three choices: the variation or intervention applied to the source variable, the response measured in the target variable, and the procedure used to aggregate responses over initial conditions or interventions.
}

\revone{
More formally, let $X$ and $Y$ denote target and source variables, respectively. Consider reference and alternative initial distributions that have the same marginal distribution of $X$ but differ in the distribution of $Y$ conditional on $X$. After both distributions are propagated by the same dynamics to a horizon $\tau$, an elementary DCE may be expressed as
\begin{equation}
    \mathcal{C}_{Y\rightarrow X}^{\tau}
    =
    \mathcal{D}
    \left(
        p_{X,\tau}^{\mathrm{alt}},
        p_{X,\tau}^{\mathrm{ref}}
    \right),
\end{equation}
where $\mathcal{D}$ is a distinction functional measuring the resulting change in the future distribution of $X$. This notation describes an elementary effect at a fixed time horizon which will be most relevant to the present work.  More generally, the DCE framework also permits a distributed temporal horizon, corresponding to a set of values of $\tau$. For example, spectral causality quantifiers have been formulated as sets of DCEs over a distributed temporal horizon \citep{Smirnov2019}. 
}

\revone{
LOCA corresponds to the deterministic, local limit of this construction. Deterministic dynamics may be regarded as the zero-noise limit of a stochastic dynamical system. For dynamics governed by Eq.~\ref{eq:g}, 
consider two initial states differing only by a perturbation $\Delta u_j$ in the source coordinate, $\boldsymbol{e}_j$. Taking the normalized change in the target coordinate (with index $i$) as the distinction functional gives
\begin{equation}
    \frac{
        g_i^{\tau}
        \left(
            \boldsymbol{u}_0+\Delta u_j\boldsymbol{e}_j
        \right)
        -
        g_i^{\tau}(\boldsymbol{u}_0)
    }{\Delta u_j}.
\end{equation}
In the infinitesimal limit, this DCE becomes
\begin{equation}
    G_{j\rightarrow i}^{\tau}
    =
    \frac{\partial g_i^{\tau}(\boldsymbol{u}_0)}
         {\partial u_j^0}.
\end{equation}
For linear dynamics, $\boldsymbol{u}_{\tau}=e^{A\tau}\boldsymbol{u}_0$, and hence
\begin{equation}
    G_{j\rightarrow i}^{\tau}
    =
    \left[e^{A\tau}\right]_{ij}.
\end{equation}
Thus, LOCA is a DCE based on an infinitesimal, pointwise intervention and a local state response. Its distinguishing feature is that, for linear or linearized systems, the causal effect is obtained directly from the evolution operator. The variance-weighted LOCA measure introduced above can similarly be viewed as a quadratic aggregation of these elementary responses, with equivalence to prediction-based measures requiring the additional assumptions to be discussed in \ref{sec:granger}.
}

\subsection{Connection with Granger causality and transfer entropy}
\label{sec:granger}

Here, we first discuss the connection between our operator-theoretic causality framework and the widely used statistical approach of Granger causality, \revone{while making clear the distinctions between the two
quantities.}

We focus on pairwise conditional Granger causality from state $c_j$ to state $c_i$ in an $m$-dimensional discrete-time system with a single time lag. Let the system state at time steps $t_n=n\,\Delta t$ be $\bc=\left[c_1(t_n),\dots,c_m(t_n)\right]^T$. Here, we use $\bc$ to denote the variables under consideration, given that in practice, for high-dimensional fluid flows, one must typically use a low-dimensional representation of the full system state, e.g., mapping the full state to POD coordinates via Eq.~\eqref{eq:xtoc}.

We will focus on pairwise conditional Granger analysis with a single time
lag, which quantifies the causality from one variable (here taken to be
$c_j$) to another ($c_i$), assuming knowledge of all other state
variables. \revone{
Note that while the original formulation and typical applications of Granger causality analysis consider multiple time lags, a single time lag will be the most appropriate point of comparison to LOCA. This is because if full state information is assumed, only a single timestep is needed to predict the future state of the deterministic system.
 For reduced or partially observed systems, however, a single time lag need not contain the complete predictive history.}

Granger causality analysis assesses causality by comparing models fitted with and without knowledge of a specified variable, \revone{termed the
unrestricted and restricted models, respectively}. The unrestricted model
for state variable $c_i(t_n)$ is
\begin{equation}
\label{eq:unrestricted}
c_i(t_n)
=
\sum_{k=1}^{m} A^g_{ik}\,c_k(t_{n-1})
+
\epsilon^u_i(t_n),
\end{equation}
where $A^g_{ik}$ are coefficients representing the influence of $c_k$ at the previous time step on the current value of $c_i$, and $\epsilon^u_i(t_n)$ is the unrestricted-model residual. The restricted model omits variable $c_j$:
\begin{equation}
\label{eq:restricted}
c_i(t_n)
=
\sum_{\substack{k=1\\k\neq j}}^{m}
A^{g,r}_{ik}\,c_k(t_{n-1})
+
\epsilon^r_{i,j}(t_n).
\end{equation}
\revone{In standard Granger analysis, the restricted model is refitted after $c_j$ is removed, and therefore the coefficients $A^{g,r}_{ik}$ need not equal the corresponding unrestricted coefficients $A^g_{ik}$.}

If excluding $c_j$ significantly degrades the prediction of $c_i$, we conclude that $c_j$ Granger-causes $c_i$. To evaluate the extent of this causality, the residual sums of squares for the restricted and unrestricted models are defined as
\begin{equation}
\mathrm{RSS}^r
=
\sum_n\left[\epsilon^r_{i,j}(t_n)\right]^2,
\qquad
\mathrm{RSS}^u
=
\sum_n\left[\epsilon^u_i(t_n)\right]^2.
\end{equation}
These quantities are compared through the $F$-statistic
\begin{equation}
\label{eq:F}
F
=
\frac{\mathrm{RSS}^r-\mathrm{RSS}^u}
{\mathrm{RSS}^u/\left(N-(m+1)\right)},
\end{equation}
where the numerator reflects the single parameter introduced by $c_j$, $N$ is the total number of data points, and the $(m+1)$ term in the denominator represents the total parameter count in the unrestricted model.

We now connect this statistical approach to our dynamical-systems framework. Suppose that in continuous time the system is locally governed by
\begin{equation}
\dot{\bc}(t)=\mathbf{A}\bc(t),
\end{equation}
so that over one time step,
\begin{equation}
\bc(t_n)
=
\mathrm{e}^{\mathbf{A}\Delta t}\bc(t_{n-1}).
\end{equation}
The unrestricted regression coefficients then correspond to elements of the discrete-time evolution operator,
\begin{equation}
A^g_{ik}
=
[\mathrm{e}^{\mathbf{A}\Delta t}]_{ik}.
\end{equation}

\revone{When the restricted model is refitted, the part of $c_j$ that is linearly predictable from the remaining variables can be absorbed into the modified coefficients $A^{g,r}_{ik}$. Consequently, only the part of $c_j$ that cannot be linearly predicted from the other retained variables contributes to the increase in restricted-model residual
variance. We denote the variance of this unpredictable component by}
\begin{equation}
\revone{
\sigma_{c_j\mid\bc_{-j}}^2,
}
\end{equation}
\revone{where $\bc_{-j}$ denotes the state vector with $c_j$ omitted. For jointly Gaussian variables, this is the conditional variance $\operatorname{var}(c_j\mid\bc_{-j})$; more generally, it is the residual variance from the best linear prediction of $c_j$ using the other variables. Assuming that the unrestricted residual is uncorrelated with the lagged state, the increase in residual variance is therefore}
\begin{equation}
\revone{
\frac{\mathrm{RSS}^r-\mathrm{RSS}^u}{N}
\approx
\left|
[\mathrm{e}^{\mathbf{A}\Delta t}]_{ij}
\right|^2
\sigma_{c_j\mid\bc_{-j}}^2.
}
\label{eq:rss-difference-granger}
\end{equation}
\revone{Under stationarity and for sufficiently large $N$, the resulting single-lag conditional Granger $F$-statistic is}
\begin{equation}
\revone{
F^{\mathrm{GC}}_{j\to i}
\approx
\frac{
\left|
[\mathrm{e}^{\mathbf{A}\Delta t}]_{ij}
\right|^2
\sigma_{c_j\mid\bc_{-j}}^2
\left(N-(m+1)\right)
}{
\sigma_{\epsilon,i}^2
}.
}
\label{eq:granger-general}
\end{equation}
\revone{Equation~\eqref{eq:granger-general} should be distinguished from the variance-weighted LOCA quantity defined in Eq.~\eqref{eq:F1}. Granger causality uses the conditional linear prediction-error variance $\sigma_{c_j\mid\bc_{-j}}^2$, whereas LOCA uses the unconditional source variance $\sigma_{c_j}^2$. A fixed-coefficient variant recovers Eq.~\eqref{eq:F1} (after removing the sample-size normalization), when the retained coefficients are held at their operator values rather than refitted after $c_j$ is removed. We omit the details, since the derivation parallels the one above. Related constructions are also discussed in \citet{Smirnov2022}. Thus, the two measures coincide when the conditional and unconditional source variances are equal, subject to the residual assumptions stated above. More generally, Granger causality measures the unique predictive contribution of $c_j$ after conditioning on the remaining variables, whereas LOCA measures the response to an independently specified variation in $c_j$.}

It is also important to note that in practical fluid applications, the residual $\epsilon_i$ often represents nonlinear effects, which may be correlated with the state variables. \revone{If the evolution operator is imposed from a linearization rather than identified as the least-squares predictor, the required orthogonality between the residual
and the lagged state need not hold. Additional covariance terms may then enter the residual-variance calculation, and the relationship between the Granger and LOCA expressions described above need not hold.}

Since transfer entropy and Granger causality are equivalent for linear Gaussian processes~\citep{Barnett2012TransferEA}, \revone{an analogous distinction applies to transfer entropy. Standard transfer entropy may use the available histories of the variables, whereas the single-lag, or strongly truncated, transfer entropy considered here conditions only on the immediately preceding state. These quantities coincide for a
complete first-order Markov state, but need not coincide for reduced or partially observed variables.}

\revone{Motivated by the logarithmic form used for Gaussian transfer entropy and Granger causality, we define the corresponding logarithmic LOCA quantity}
\begin{equation}
\revone{
F^{(L)}_{j\to i}
=
\log\left(
1+
\frac{
\left|
[\mathrm{e}^{\mathbf{A}\Delta t}]_{ij}
\right|^2
\sigma_{c_j}^2
}{
\sigma_{\epsilon,i}^2
}
\right).
}
\label{eq:log-loca}
\end{equation}
\revone{This is a logarithmic transformation of the LOCA statistic in Eq.~\eqref{eq:F1}, rather than transfer entropy in general. It coincides with the corresponding single-lag Gaussian Granger-causality or transfer-entropy expression when the conditional and unconditional source variances are equal and the relevant residual variances are identified.} In some works, an analogous logarithmic quantity is also used within the context of Granger analysis
\citep{barnett2014mvgc,barnett2015granger}.

\revone{Both Granger causality and transfer entropy can also be characterized as particular constructions within the DCE framework \citep{SmirnovMokhov2015,Smirnov2020}. LOCA, Granger causality, and transfer entropy therefore correspond to different choices of variation and response functional within this broader framework, and should not be
expected to agree outside the specific conditions identified above.}

\subsection{Connection with graph theory \revone{and derivation of global causality indicator}}
\label{sec:graph}
In this section, we will show how notions from graph theory can be related both mathematical and conceptually to the notions of causality discussed in Sect.~\ref{sec:causality}. Once again, we consider a linear system with dynamics $\dot\bc = \mathbf{A} \bc$, with $A_{ij}$ denoting the elements of $\mathbf{A}$. 

We first note that even if $A_{ij} = 0$, indicating no direct coupling from mode $j$ to mode $i$ at any given instance, indirect coupling can arise through interactions with intermediate modes. This indirect influence becomes evident over longer time horizons due to the cumulative effect of the system's interconnectedness. This becomes explicit in the power series expansion of the matrix exponential:
\begin{equation}\label{eq:IndirectCausality}
\mathrm{e}^{\mathbf{A} t} = \mathbf{I} + \mathbf{A} t + \frac{1}{2} \mathbf{A}^2 t^2 + \frac{1}{6} \mathbf{A}^3 t^3 + \dots
\end{equation}
Each term $\mathbf{A}^k t^k / k!$ in this series represents the cumulative influence of interactions over time. Importantly, higher-order terms $\mathbf{A}^k$ for $k \geq 2$ describe the indirect coupling pathways through which modes can influence one another via intermediate modes. If $A_{ij} = 0$, there is no direct influence of mode $j$ on mode $i$. However, consider the case where mode $j$ influences mode $k$ ($A_{kj} \neq 0$), and mode $k$ influences mode $i$ ($A_{ik} \neq 0$). This creates an indirect pathway from mode $j$ to mode $i$, which is captured by the second-order term $\mathbf{A}^2$. Specifically:
\begin{equation}
(\mathbf{A}^2)_{ij} = \sum_{l} A_{il} A_{lj}
\end{equation}
This summation includes all possible intermediate modes $l$ through which mode $j$ can indirectly influence mode $i$ and includes the non-zero term $A_{ik} A_{kj}$. Higher-order terms $\mathbf{A}^3$, $\mathbf{A}^4$, etc., describe indirect interactions through longer chains of intermediate modes, progressively accounting for more complex indirect pathways. For smaller time horizons $t\ll1$, the higher order powers of $\mathbf{A}$ may be ignored, in which case, there is no causal connection between modes $i$ and $j$. As $t$ increases, these higher-order terms can no longer be ignored. Over longer time horizons, the contributions of higher-order terms in the matrix exponential become more significant, especially in systems with strong interactions between modes. Thus, even when there is no direct coupling ($A_{ij} = 0$), indirect effects will become observable through the cumulative influence of intermediate pathways. 

This picture can be viewed from the perspective of graph theory where the nodes are the coefficients $c_i$ and the $\mathbf{A}$ matrix is related to the adjacency matrix. To compare analogous concepts in graph theory, we will denote the adjacency matrix with the same symbol $\mathbf{A}$. In graph theory, an \textit{adjacency matrix} $\mathbf{A}$ is a square matrix used to represent a finite graph \citep{west2001introduction}. The elements of the matrix $\mathbf{A}$, denoted $A_{ij}$, describe the connections between nodes (or vertices) of the graph. Specifically, $A_{ij} = 1$ if there is a direct edge from node $i$ to node $j$, and $A_{ij} = 0$ otherwise. The adjacency matrix is symmetric for an undirected graph, meaning $A_{ij} = A_{ji}$ for all $i,j$. In the case of directed graphs, the matrix is generally asymmetric, with $A_{ij} = 1$ only if there is a directed edge from $i$ to $j$. 

The concept of \textit{transitive closure} (first introduced in relational mathematics and set theory \citep{tarski1941calculus}) of a graph is closely related to the adjacency matrix. The transitive closure of a graph captures not only the direct connections between nodes but also the indirect connections that can be reached through a sequence of edges. 
It is given by the Boolean Neumann series \citep{Warshall1962ATO}:
\begin{equation}\label{eq:transitive}
\mathbf{T} = \mathbf{I} \lor \mathbf{A} \lor \mathbf{A}^2 \lor \mathbf{A}^3 \lor \cdots = \bigvee_{k=0}^{\infty} \mathbf{A}^k.
\end{equation}

\revone{We use this graph-theoretic analogy as a template for defining
a cancellation-free global indicator of causal pathways. The successive
powers of $\mathbf{A}$ encode hierarchies of indirect interactions, but
a direct sum such as $\sum_k \mathbf{A}^k$ can include cancellations
between positive and negative, or more generally complex-valued,
contributions. To avoid such cancellations, we instead use the
elementwise absolute value. Since}
\begin{equation}
\revone{
    |\mathbf{A}^k|
    \leq
    |\mathbf{A}|^k
}
\end{equation}
\revone{elementwise, the powers of $|\mathbf{A}|$ provide an upper bound on the magnitude of all length-$k$ pathways.}

\revone{This construction is also directly connected to the finite-time LOCA response. From the power-series expansion of the matrix exponential,
\begin{align}
    |\mathrm{e}^{\mathbf{A}t}|
    &=
    \left|
    \sum_{k=0}^{\infty}
    \frac{\mathbf{A}^k t^k}{k!}
    \right|
    \leq
    \sum_{k=0}^{\infty}
    \frac{|\mathbf{A}^k|t^k}{k!} \\
    &\leq
    \sum_{k=0}^{\infty}
    \frac{|\mathbf{A}|^k t^k}{k!}
    =
    \mathrm{e}^{|\mathbf{A}|t}.
\end{align}
Thus, $\mathrm{e}^{|\mathbf{A}|t}$ is a pointwise, cancellation-free envelope for the magnitude of the causal response at each fixed temporal horizon. However, this bound remains time-dependent and does not by itself provide a finite global summary over all future times. We therefore introduce the exponentially discounted cumulative absolute response
\begin{equation}
    \mathbf{C}_\alpha
    =
    \int_0^\infty
    \mathrm{e}^{-\alpha t}
    |\mathrm{e}^{\mathbf{A}t}|\,\mathrm{d}t,
    \qquad
    \alpha>\rho(|\mathbf{A}|).
\end{equation}
Using the pointwise envelope above,
\begin{align}
    \mathbf{C}_\alpha
    &\leq
    \int_0^\infty
    \mathrm{e}^{-\alpha t}
    \mathrm{e}^{|\mathbf{A}|t}\,\mathrm{d}t \\
    &=
    \int_0^\infty
    \mathrm{e}^{(|\mathbf{A}|-\alpha\mathbf{I})t}\,\mathrm{d}t \\
    &=
    (\alpha\mathbf{I}-|\mathbf{A}|)^{-1}
    :=
    \mM_\alpha.
    \label{eq:M}
\end{align}
The condition $\alpha>\rho(|\mathbf{A}|)$ ensures that the integral converges. Thus, $\mM_\alpha$ is an upper bound on the exponentially discounted, time-integrated magnitude of the finite-time causal response. Equivalently,
\begin{equation}
    \mM_\alpha
    =
    \frac{1}{\alpha}
    \left(
        \mathbf{I}
        -
        \frac{|\mathbf{A}|}{\alpha}
    \right)^{-1}
    =
    \sum_{k=0}^{\infty}
    \frac{|\mathbf{A}|^k}{\alpha^{k+1}},
\end{equation}
so $\mM_\alpha$ can also be interpreted as a geometrically weighted, cancellation-free sum over interaction pathways of increasing length.}

\revone{The zero pattern of $\mM_\alpha$ provides a rigorous criterion for causal disconnection. Because every term in the Neumann series above is elementwise nonnegative, $[\mM_\alpha]_{ij}=0$ implies $[|\mathbf{A}|^k]_{ij}=0$ for every $k\geq 0$. Since $|[\mathbf{A}^k]_{ij}|\leq [|\mathbf{A}|^k]_{ij}$, it follows that $[\mathbf{A}^k]_{ij}=0$ for every $k\geq 0$, and therefore
\begin{equation}
    [\mathrm{e}^{\mathbf{A}t}]_{ij}
    =
    \sum_{k=0}^{\infty}
    \frac{[\mathbf{A}^k]_{ij}t^k}{k!}
    =
    0
    \qquad
    \text{for all }t\geq 0.
\end{equation}
Thus, a zero entry of $\mM_\alpha$ certifies that state $j$ has no causal influence on state $i$ at any temporal horizon. A positive entry, however, should be interpreted as indicating the presence of one or more structural interaction pathways, rather than as a lower bound on the response at every time. The actual finite-time response may still be reduced or vanish at particular times because of sign or phase cancellations.}

\revone{For visualization and comparison across systems, it can also be useful to remove the dimensional dependence on the choice of time units. This can be achieved by setting
\begin{equation}
    \alpha
    =
    \gamma\rho(|\mathbf{A}|),
    \qquad
    \gamma>1,
\end{equation}
and define the dimensionless matrix
\begin{equation}
    \mathbf{K}_\gamma
    =
    \alpha\mM_\alpha
    =
    \left(
        \mathbf{I}
        -
        \frac{|\mathbf{A}|}{\gamma\rho(|\mathbf{A}|)}
    \right)^{-1}.
\end{equation}
This normalization is invariant under a uniform rescaling of time. In particular, if $\mathbf{A}$ is replaced by $\beta\mathbf{A}$ with $\beta>0$, then both $|\mathbf{A}|$ and $\rho(|\mathbf{A}|)$ are multiplied by $\beta$, leaving $\mathbf{K}_\gamma$ unchanged. In the numerical results below, we use $\gamma=1.05$, which places $\alpha$ five percent above the convergence threshold. Larger values of $\gamma$ suppress longer pathways more strongly, while values closer to one give greater relative weight to longer pathways but lead to a more ill-conditioned inverse.}

\revone{If $\rho(|\mathbf{A}|)=0$, such as for a strictly triangular
$\mathbf{A}$ whose elementwise absolute value is nilpotent, the choice $\alpha=\gamma\rho(|\mathbf{A}|)$ is not applicable. In that case, any
prescribed $\alpha>0$ yields a finite series, or $\alpha$ may be chosen using an independent characteristic inverse timescale.}

\revone{The matrix $\mM_\alpha$ also has a formal connection to the resolvent, since $\int_0^\infty \mathrm{e}^{-\alpha t}\mathrm{e}^{\mathbf{A}t} \,\mathrm{d}t=(\alpha\mathbf{I}-\mathbf{A})^{-1}$. Thus, $\mM_\alpha$ may be viewed as a cancellation-free upper bound on a discounted resolvent response. It also bears formal similarity to centrality measures in graph theory, especially Katz centrality \citep{katz1953new}, but here it is used to bound cumulative causal influence and to identify causally disconnected state pairs.}

\subsection{Connection with observability and controllability}
\label{sec:bt}
Augmenting the linear system under consideration with a control input ($\mathbf{u}$) and an output ($\mathbf{y}$), we have:
\begin{equation}
\dot{\bc } = \mathbf{A}\bc  + \mathbf{B}\mathbf{u}, 
\end{equation}
\begin{equation}
\mathbf{y} = \mathbf{C}\bc .
\end{equation}
The \emph{controllability matrix} for this system is defined as
\begin{equation}
\mathcal{C} = \begin{bmatrix} \mathbf{B} & \mathbf{A}\mathbf{B} & \mathbf{A}^2\mathbf{B} & \cdots & \mathbf{A}^{n-1}\mathbf{B} \end{bmatrix},
\end{equation}
and the \emph{observability matrix} is given by
\begin{equation}
\mathcal{O} = \begin{bmatrix} \mathbf{C} \\[4mm] \mathbf{C}\mathbf{A} \\[2mm] \mathbf{C}\mathbf{A}^2 \\[2mm] \vdots \\[2mm] \mathbf{C}\mathbf{A}^{n-1} \end{bmatrix}.
\end{equation}
Suppose that $M_{ij}=0$ for some indices $i\neq j$. As discussed earlier, this implies:
\begin{equation}
\left[ \mathbf{A}^k \right]_{ij} = 0, \quad \forall\, k \ge 0.
\end{equation}
Assume further that the control input is applied only to the $j^{\mathrm{th}}$ state; that is, only $B_j\neq 0$. Then, the $i^{\mathrm{th}}$ component of each term in the controllability matrix is given by
\begin{equation}
\left[ \mathbf{A}^k\mathbf{B} \right]_i = \sum_{l=1}^n \left[ \mathbf{A}^k \right]_{il} B_l 
= \left[ \mathbf{A}^k \right]_{ij} B_j = 0 \quad \forall\, k.
\end{equation}
This means that the entire $i^{\mathrm{th}}$ row of $\mathcal{C}$ is zero, implying that $\mathcal{C}$ is rank deficient and the system is not fully controllable. Thus, $M_{ij}=0$ results in the loss of controllability in the system if the control input is only applied to the $j^{\mathrm{th}}$ mode. Similarly, assume that the measurement is taken solely from the $i^{\mathrm{th}}$ state so that
\begin{equation}
\mathbf{C} = \begin{bmatrix} 0 & \cdots & C_i & \cdots & 0 \end{bmatrix}.
\end{equation}
Then, for every $k$ we have
\begin{equation}
\mathbf{C}\mathbf{A}^k = C_i \left[ \mathbf{A}^k \right]_{ij} = 0,
\end{equation}
which shows that the contribution from the $j^{\mathrm{th}}$ state to the output is never observed. Consequently, the observability matrix $\mathcal{O}$ loses rank, and the $j^{\mathrm{th}}$ state is unobservable. This connection between causality and controllability/observability also extends to the Gramians, which are defined by:
\begin{equation}
    \mathbf{W}_o = \int_0^\infty \mathrm{e}^{\mathbf{A}^*\tau}\mathrm{e}^{\mathbf{A}\tau} \mathrm{d}t,
\end{equation}
\begin{equation}
    \mathbf{W}_c = \int_0^\infty \mathrm{e}^{\mathbf{A}\tau}\mathrm{e}^{\mathbf{A}^*\tau} \mathrm{d}t,
\end{equation}

These Gramians can also be used to find reduced-order models that optimally retain observability and controllability of the system via balanced truncation \citep{moore1981principal}. In particular, here state truncation is performed after first transforming the coordinates such that $\mathbf{W}_o$ and $\mathbf{W}_c$ are diagonal and identical, with ordered diagonal entries corresponding to the Hankel singular values of the system. 
For high-dimensional  systems, this can be implemented without needing to explicitly form the associated operators using forward and adjoint simulations through balanced proper orthogonal decomposition \citep{rowley:05pod,willcox:2002}. We refer the reader to these references for further details regarding these methods.

It is well known that if the controllability and observability matrices are not full rank, then the corresponding Gramians are singular. Therefore, there is a direct connection between the condition $M_{ij}=0$ and the singularity of the Gramians. This has the interesting consequence that the metric $\mM_\alpha$ can serve as an alternative to detecting controllability and observability, and the elements of $\mM_\alpha$ could be useful for guiding appropriate locations for sensor and actuator placement.

\section{Causal inference from data}\label{sec:causal-inference}

Since causality is embedded in the system matrix $\mathbf{A}$, it is of practical relevant to determine if the elements of this matrix can be ascertained from data. 
Here, we propose one method for doing this, based on the dynamic mode decomposition (DMD) \citep{Schmid2008DynamicMD} technique. DMD is a data-driven technique used to approximate the underlying dynamics of a complex system from time-series data. In the context of a linear dynamical system, DMD allows us to estimate the system matrix, which can then be used to infer causal relationships between the components.

Let $\mathbf{X}$ 
denote a data matrix, where each column of $\mathbf{X}$ represents the state vector at a specific time step. We define two data matrices,
\begin{equation}
    \mathbf{X}_1 = [\bc (t_1), \bc (t_2), \ldots, \bc (t_{N-1})],
    \quad
   \mathbf{X}_2 = [\bc (t_1+\Delta t), \bc (t_2+\Delta t), \ldots, \bc (t_{N-1}+\Delta t)],
\end{equation}
which capture the system's state at consecutive time steps. If using a single initial condition with $N$ timesteps of data, then $\mathbf{X}_1$ and $\mathbf{X}_2$ consist of $\mathbf{X}$ with the last and first column removed, respectively. The goal of DMD here is to find an approximate linear map $\mathbf{A}_{\mathrm{DMD}}$ such that
\begin{equation}
    \mathbf{X}_2 \approx \mathbf{A}_{\mathrm{DMD}} \mathbf{X}_1.
\end{equation}
To solve for $\mathbf{A}_{\mathrm{DMD}}$, we use the least-squares solution given by
\begin{equation}
\label{eq:DMD}
    \mathbf{A}_{\mathrm{DMD}} = \mathbf{X}_2 \mathbf{X}_1^{\dagger},
\end{equation}
where $\mathbf{X}_1^{\dagger}$ denotes the Moore-Penrose pseudoinverse of $\mathbf{X}_1$ \citep{tu2014dynamic}. This approach yields the matrix $\mathbf{A}_{\mathrm{DMD}}$ that best approximates the linear mapping from $\mathbf{X}_1$ to $\mathbf{X}_2$ in a least-squares sense. This estimated matrix $\mathbf{A}_{\mathrm{DMD}}$ gives us a relation between $\bc(t_{n})$ and $\bc(t_{n+1})$.
Accordingly, \(\mathbf{A}_{\mathrm{est}} = (\mathbf{A}_{\mathrm{DMD}} - \mathbf{I})/\Delta t\) provides a first-order estimate of the continuous-time system matrix $\mathbf{A}$ associated with the underlying dynamics, with a higher-order estimate given by
$\log_e(\mathbf{A}_{\mathrm{DMD}})/\Delta t$. 
We can use the elements of $\mathbf{A}_{\mathrm{est}}$ to now perform causality analysis by computing $\mM_\alpha=(\alpha\mathbf{I}-|\mathbf{A}_{\mathrm{est}}|)^{-1}$, where $\alpha$ is taken to be $1.05\rho(|\mathbf{A}_{\mathrm{est}}|)$.

\begin{figure}
    \centering
    \includegraphics[width=0.7\textwidth]{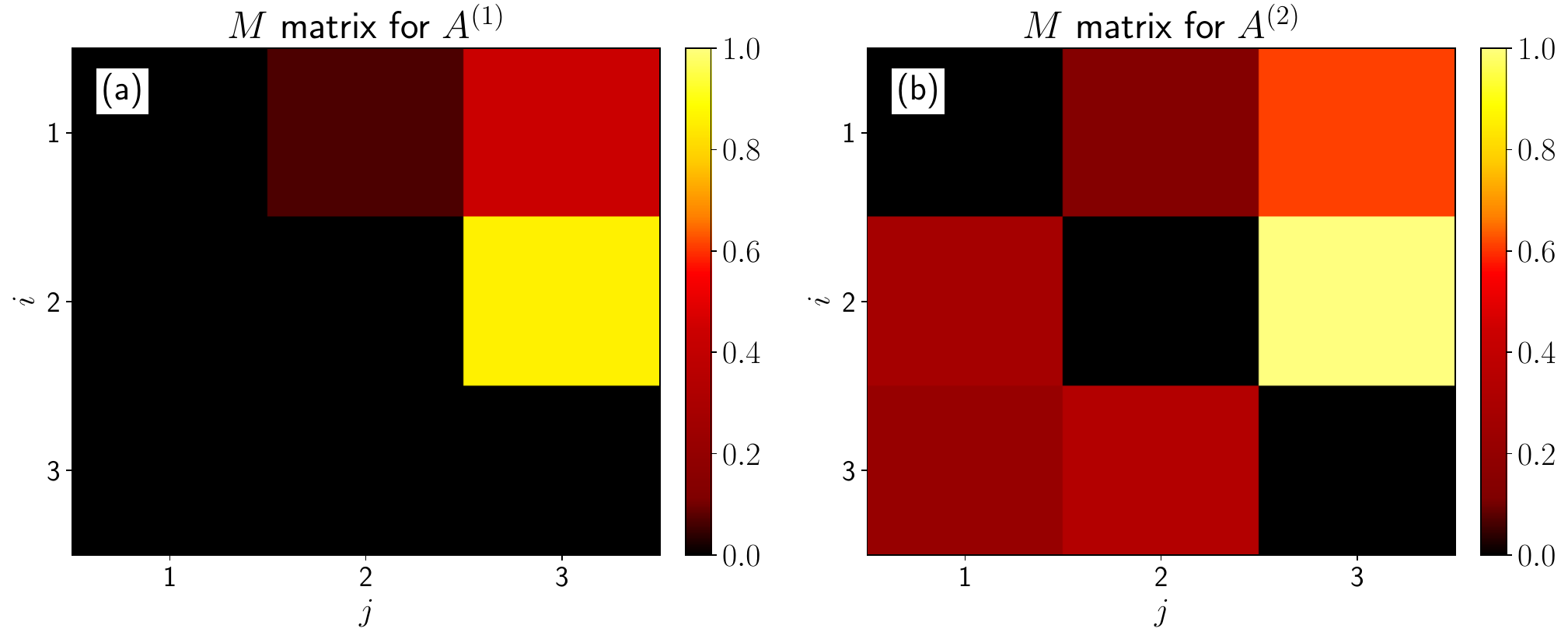}
    \caption{Global causality estimates for the two systems given in Eq.~\eqref{eq:exampleA}. (a) The system matrix ($\mathbf{A}^{(1)}$) is upper triangular and the corresponding $\mM_\alpha$ is consequently also upper triangular. (b) Immediate causality in $\mathbf{A}^{(2)}$ from $1\rightarrow 2$ and $2\rightarrow 3$ is zero, but causal connections emerge in $\mM_\alpha$ along these directions through indirect channels.}
    \label{fig:causality_matrices}
\end{figure}

To illustrate the process, we consider specific systems governed by a different $\mathbf{A}$ matrix. We then generate data $\bc ^{(i)}(t)$ based on random initial conditions $\bc ^{(i)}_0$ for $i=1,2,3,\cdots,N_i$. In the following examples, we generally take $N_i\approx 500$ different initial conditions, and for each initial condition, we simulate for 100 time units with $\Delta t=0.1$). This data, in totality, is then used to calculate $\mathbf{A}_{\mathrm{DMD}}$ via Eq.~\eqref{eq:DMD} and then $\mathbf{A}_{\mathrm{est}}$ and $\mM_\alpha$. Causality matrices are then presented based upon normalized values of $|\mM_\alpha|$. The matrices considered are:
\begin{align}\label{eq:exampleA}
\mathbf{A}^{(1)} = 
\begin{bmatrix}
-1 & 1 & 0 \\
0 & -2 & 1 \\
0 & 0 & -3
\end{bmatrix}
\quad \text{and} \quad
\mathbf{A}^{(2)} = 
\begin{bmatrix}
-1 & 1 & 0 \\
0 & -2 & 1 \\
0.5 & 0 & -3
\end{bmatrix}
\end{align}

These results for two different stable $\mathbf{A}$ matrices are presented in Fig.~\ref{fig:causality_matrices}. Figure \ref{fig:causality_matrices}(a) corresponds to an upper triangular $\mathbf{A}^{(1)}$ matrix with eigenvalues $\lambda =-1,-2,-3$. The DMD estimates  for both $\mathbf{A}^{(1)}$ and $\mathbf{A}^{(2)}$ results in identified matrices with very similar eigenvalues, with a normed error less than $0.002$ in both cases. The identified $\mM_\alpha$ matrix corresponding to $\mathbf{A}^{(1)}$ is also upper triangular, showing no causal flow from $1\rightarrow 2,1\rightarrow 3,2\rightarrow 3$. Figure  \ref{fig:causality_matrices}(b) corresponds to an upper triangular $\mathbf{A}^{(2)}$ matrix with a nonzero term in the bottom left corner. 
It is interesting to note that even though $A^{(2)}_{21}=A^{(2)}_{32}=0$, $M^{(2)}_{21},M^{(2)}_{32}\neq0$ indicating indirect channels of global causality. Although not shown here, we have tested our framework for larger systems, where it emerges as a highly data-efficient technique for gauging causality.

\section{Application: linearized Couette flow}
\label{sec:resultsCouette}
We now test the performance and behavior of the proposed causality analysis methodology on a fluid flow example. We consider incompressible flow between two parallel plates of infinite extent, with the flow driven by relative plate motion.
We assume the plates are in the $x$--$z$ plane, at $y= \pm 1$, with plate motions $U(y=\pm 1)=\pm 1$. The laminar solution gives the classic linear planar Couette velocity profile in the $x$-direction, given by
\begin{equation}
    U(y) = y.
\end{equation}
 While we focus exclusively on this profile, we find that other channel flow velocity profiles (e.g.~planar Poiseuille flow) produce comparable results. 
The equations of motion linearized about this base velocity profile can be expressed in terms of the wall-normal velocity ($v$) and vorticity ($\eta$). Homogeneity in the streamwise ($x$) and spanwise ($z$) directions allow for the use of Fourier transforms in these directions, such that we can independently consider Fourier modes of the form
\begin{equation}
v(x,y,z,t) = \hat{v}(y,t)\mathrm{e}^{\mathrm{i} k_x x+\mathrm{i}k_z z}
\end{equation}
\begin{equation}
\eta(x,y,z,t) = \hat{\eta}(y,t)\mathrm{e}^{\mathrm{i} k_x x+\mathrm{i}k_z z},
\end{equation}
where $k_x$ and $k_z$ are the streamwise and spanwise wavenumbers. The linearized Navier--Stokes equations can be expressed as
\begin{equation}
\label{eq:chan}
  \mathbf{E}  \frac{\partial}{\partial t} \begin{pmatrix}
\hat{v} \\
    \hat\eta
    \end{pmatrix}
    =
\begin{pmatrix}
\mathbf{L}_{vv} & 0 \\
\mathbf{L}_{v\eta} & \mathbf{L}_{\eta\eta}
    \end{pmatrix}
    \begin{pmatrix}
    \hat{v} \\
    \hat\eta
    \end{pmatrix} = \mathbf{L} \begin{pmatrix}
    \hat{v} \\
    \hat\eta
    \end{pmatrix},
\end{equation}
where the constituent operators are given by
\begin{equation}
    \mathbf{E} = 
    \begin{pmatrix}
    \Laplace & 0 \\
    0 & \mathbf{I}
\end{pmatrix},
\end{equation}
\begin{equation}
\mathbf{L}_{vv} =  \mathrm{i} k_x U \Laplace- \mathrm{i} k_x U_{yy} - \frac{1}{\Rey} \Laplace^2, \label{eq:os}
\end{equation}
\begin{equation}
\mathbf{L}_{\eta\eta} =  \mathrm{i} k_x U - \frac{1}{\Rey}\Laplace,
\end{equation}
\begin{equation}
\mathbf{L}_{\eta v} = \mathrm{i}k_z U_y.
\end{equation}
Here $\Rey$ is the Reynolds number (based on channel half-height and maximum baseflow velocity), and $\Laplace = \mathbf{D}_{yy}-k_x^2-k_z^2$ is the Laplace operator, with $\mathbf{D}_{yy}$ denoting the second derivative in the $y$ direction. $U_y$ and $U_{yy}$ denote the first and second derivatives of the baseflow in $y$ (though for the Couette flow case considered here $U_{yy}=0)$. The corresponding discrete-time system of equations can be expressed as 
\begin{equation}
\label{eq:chanDisc}
\begin{pmatrix}
    \hat{v} \\
    \hat\eta
    \end{pmatrix}_{k+1} = \mathbf{A}_d \begin{pmatrix}
    \hat{v} \\
    \hat\eta
    \end{pmatrix}_{k},
\end{equation}
where the discrete-time dynamics $\mathbf{A}_d$ corresponding to a timestep ($\Delta t$) are given by
\begin{equation}
\label{eq:Adchan}
    \mathbf{A}_d = \exp\left[(\Delta t)\mathbf{E}^{-1} \mathbf{L} \right].
\end{equation}
The no-slip boundary condition for the velocity perturbations, along with the continuity equation, results in Dirichlet boundary conditions for wall-normal vorticity fluctuations ($\hat\eta(y=\pm 1) = 0$), and clamped boundary conditions for wall-normal velocity fluctuations ($\hat{v}(y=\pm 1) = \mathbf{D}_y\hat{v}(y=\pm 1) = 0$). 
As an aside, note that the imposition of no-slip boundary conditions makes $\mathbf{E}$ invertible, making Eq.~\eqref{eq:Adchan} well-posed. 
We chose this velocity-vorticity formulation primarily because it clearly shows the distinct underlying physical mechanisms associated with the linearized dynamics. These equations are discretized with a Chebyshev pseudospectral collocation method. 
We analyze causality by considering the equivalent discrete-time operator for various time horizons.For the results to be shown here we fix $\Rey = 2000$ and $k_x = k_z = 1$, though we observe similar findings across a broad range of parameter values. 

\begin{figure}
 \centering {
\includegraphics[width=0.7\textwidth]{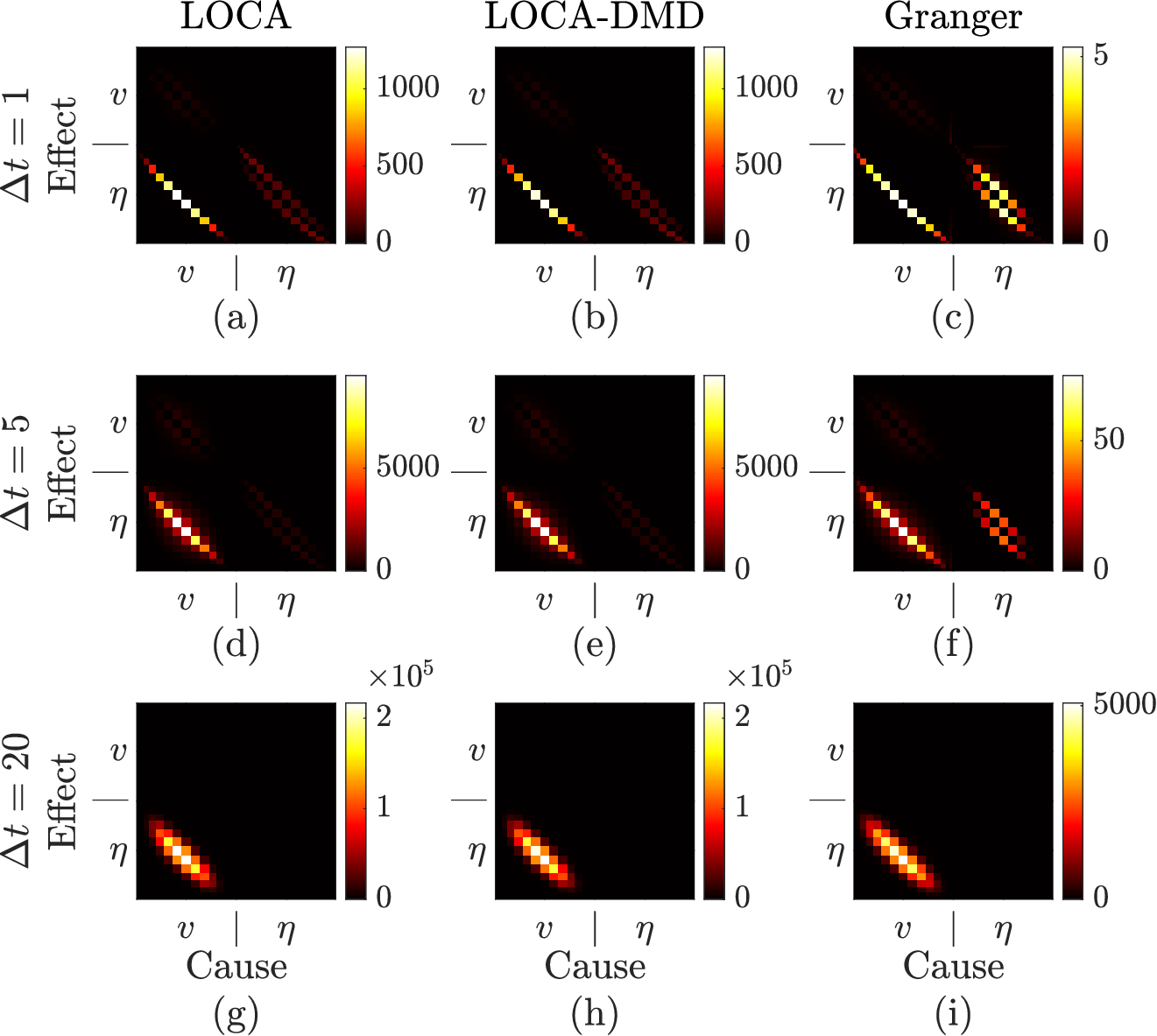}
}
\caption{Causality metrics (amplitude of $F$-statistic) obtained using LOCA (a,d,g), LOCA-DMD (b,e,h) and \revone{single-lag} Granger analysis (c,f,i) for Couette flow with $\Rey=2000$, and $k_x = k_z = 1$, over time horizons $\Delta t = 1$ (a-c), 5 (d-f) and 20 (g-i). Each pixel corresponds to $v$ or $\eta$ at a given location in the domain. Pixels in the center of the domain are larger, corresponding to the larger distance between collocation points in this region. The distinction between Granger and LOCA-based methods is evident in the structure and/or amplitude of $F$ in all cases.}
\label{fig:chan1}
\end{figure}

Figure \ref{fig:chan1} plots the conditional pairwise causality between the $v$ and $\eta$ components across all $y$ locations in the domain. We consider our proposed causality metric computed directly from the operator (LOCA) and obtained from data via DMD (LOCA-DMD) and \revone{single-lag} Granger causality computed using the same dataset. 
For the LOCA case, no data is used for the computation, with state variances computed by solving  Eq.~\eqref{eq:dlyap} with a known error covariance matrix. 
For the data-driven cases (LOCA-DMD and Granger), the data is generated by running simulations of the discrete-time system Eq.~\eqref{eq:Adchan} (equivalent to running the continuous time system with a forward Euler method), forced by uncorrelated white noise in each state of variance $\sigma_N^2 = 0.01$. 
\revone{We emphasize that, as is standard, the restricted and unrestricted Granger models are identified using separate regressions. Consequently, the Granger statistic depends on the component of the source variable that is not linearly predictable from the remaining variables, whereas LOCA uses the unconditional source variance. The conditions for equivalence discussed in Sect.~\ref{sec:granger} therefore need not hold.} 
For the purposes of this example, we adjust the timestep to match the time lag ($\Delta t$) over which causality is analyzed. This allows us to directly control the error in the discrete time system through control of  $\sigma_N^2$. In the example considered in Sect.~\ref{sec:results-wake}, we will consider a case where we do not have such control over the properties of the error/noise in the linear dynamics. The results are shown for $N_y = 16$ Chebyshev collocation points, though similar results are obtained for larger $N_y$. The data-driven methods use $N_t = 10~000$ timesteps of data. 
Following the convention for Granger-based causality analysis, we exclude computations of self-causality (which correspond to the diagonal components of the discrete-time dynamics). We can make several observations. First, we observe good agreement between the theoretical operator-based analysis, and the analysis computed empirically from data. 
\revone{Second, we observe notable differences between the LOCA and
single-lag Granger results, as may be expected because the two methods
measure different quantifiers of the source--target relationship. For small time steps, the single-lag Granger analysis assigns relatively large causality values between the $\eta$ components of adjacent elements} 
(which correspond to entries adjacent to the diagonal in the $\mathbf{L}_{\eta\eta}$ sub operator, as well as in the diagonal entries of the $\mathbf{L}_{\eta v}$).
In contrast, our causality metric identifies the diagonal of $\mathbf{L}_{\eta v}$ to be the largest causal mechanism for all timesteps. This corresponds to a fluctuation in $v$ leading to a response (causal effect) in $\eta$. This can be associated with the well-known lift-up mechanism \citep{landahl1975wave}. 
\revone{We interpret the relatively large single-lag Granger values in the $\mathbf{L}_{\eta\eta}$ sub-operator at smaller timesteps as reflecting the predictive information carried by neighboring $\eta$ signals after the restricted model is refitted. Because these signals are strongly correlated, this prediction-based measure need not rank the corresponding interactions in the same manner as the intervention-based response quantified by LOCA.} 
At larger timesteps, all causality maps converge to a similar structure, dominated by the central band of the $\mathbf{L}_{\eta v}$ operator. 

\begin{figure}
 \centering {
\includegraphics[width= 0.6\textwidth]{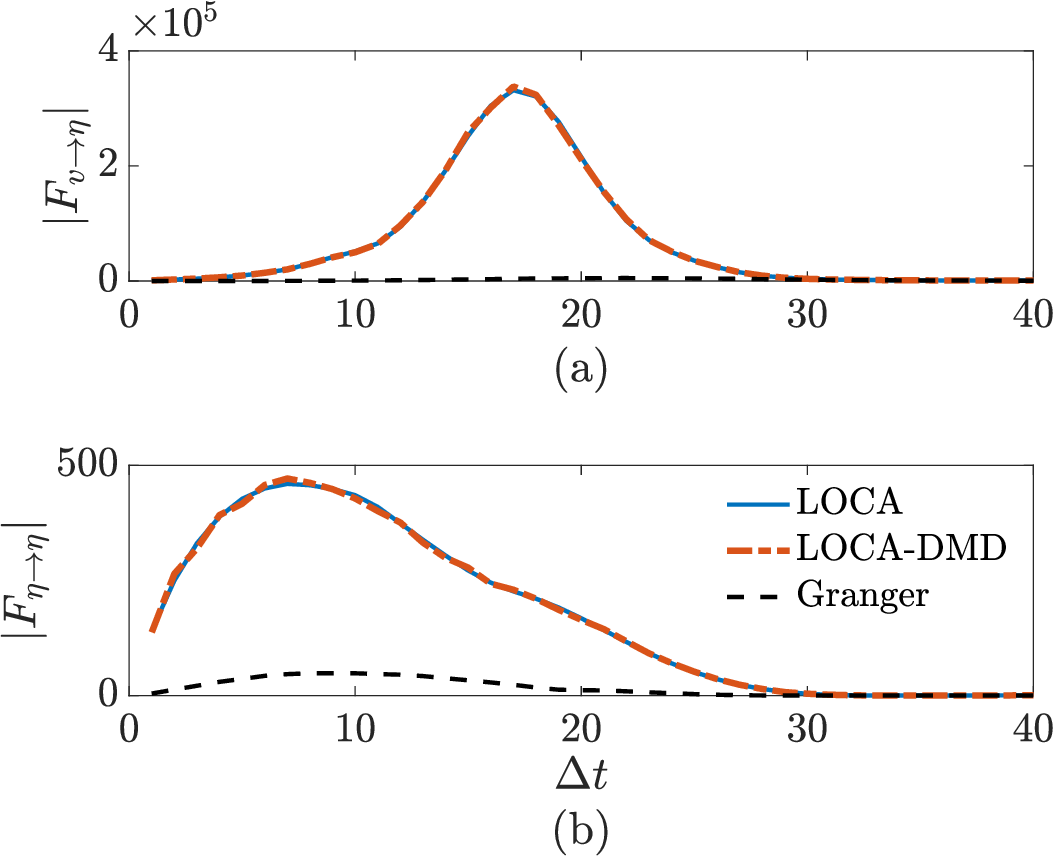}
}
\caption{Amplitude of $F$-statistic for all methods as a function of the timestep, for linearized Couette flow with the same parameters as Fig.~\ref{fig:chan1}. Cause and effect variables are taken at the same location in the center of the channel for $v\to\eta$ (a) and adjacent locations near the channel center for $\eta\to\eta$ (b).}
\label{fig:chan2}
\end{figure}

It can also be observed in Fig.~\ref{fig:chan1} that the magnitude of the $F$-statistics for our causal metric is much larger than Granger. To study this in more detail, Fig.~\ref{fig:chan2} shows the evolution of the $F$-statistic amplitude for all methods as a function of the timestep at locations near the center of the channel. Here, the distinction in amplitude between the two causal analysis methods is further apparent, as is the agreement between the LOCA and LOCA-DMD methods.

\begin{figure}
 \centering {
\includegraphics[width= 0.65\textwidth]{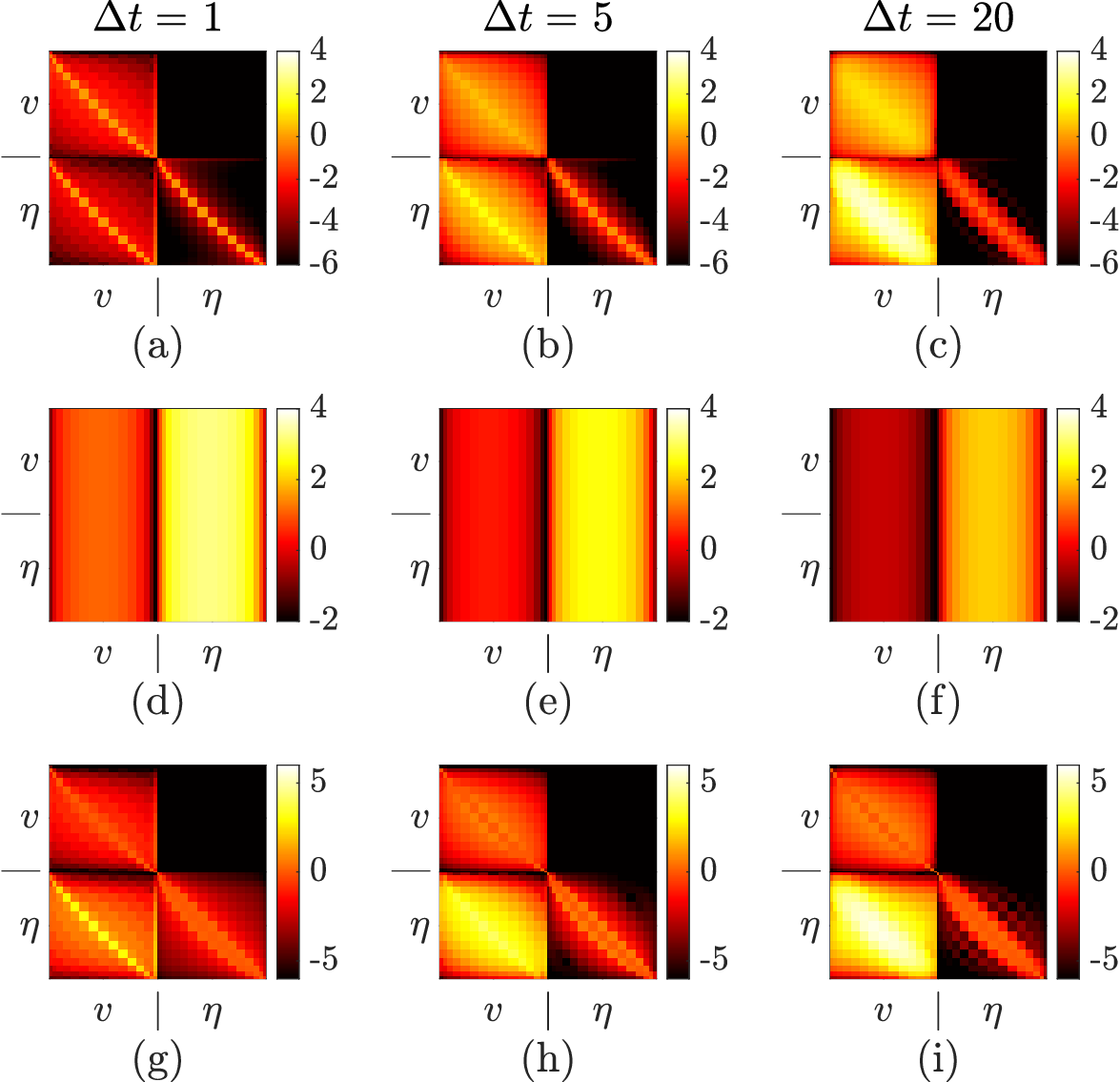}
}
\caption{ (a-c) Log amplitude of entries of discrete time operators $\log_{10}\left(\left[\mathbf{A}_d\right]_{ij}^2\right)$, (d-f) log of state variance $\log_{10}(\sigma_j^2)$, and (g-i) log of the product of these terms $\log_{10}\left(\left[\mathbf{A}_d\right]_{ij}^2\sigma_j^2\right)$, which is proportional to $F$ assuming uniform noise in each state. Quantities are computed for time horizons $\Delta t = 1$ (a-c), 5 (d-f), and 20 (g-i).}
\label{fig:chanMats}
\end{figure}

To understand the contributors to these results of LOCA-based causality analysis in more detail, in Fig.~\ref{fig:chanMats} we show the contributing terms to $F$ for LOCA, for the same time horizons considered in Fig.~\ref{fig:chan1}. In particular, assuming that the noise (error) in the linearized dynamics has the same amplitude for all states, $F_{j\to i}$ is proportional to the product $\left[\mathbf{A}_d\right]_{ij}^2\sigma_j^2$. Note in particular that the quantities plotted in Fig.~\ref{fig:chanMats}(g-i) are proportional to those in Fig.~\ref{fig:chan1}(a,d,g) but plotted on a log scale to highlight the different scales associated with each term (and the quantities in Fig.~\ref{fig:chanMats} also include the diagonal entries of the operators). Figure \ref{fig:chanMats}(a-f) shows that the large amplitudes for $F_{v\to\eta}$ are caused by the large entries in the $\mathbf{A}_d$  operator, which overcome the smaller amplitude of the $v$ states in comparison to $\eta$. In other words, even though the computation of $F_{v\to\eta}$ accounts for the smaller energy of the causing $v$ variable, the size of the terms in the corresponding entries of $\mathbf{A}_d$ overcome this. 

\begin{figure}
 \centering {
\includegraphics[width= 0.7\textwidth]{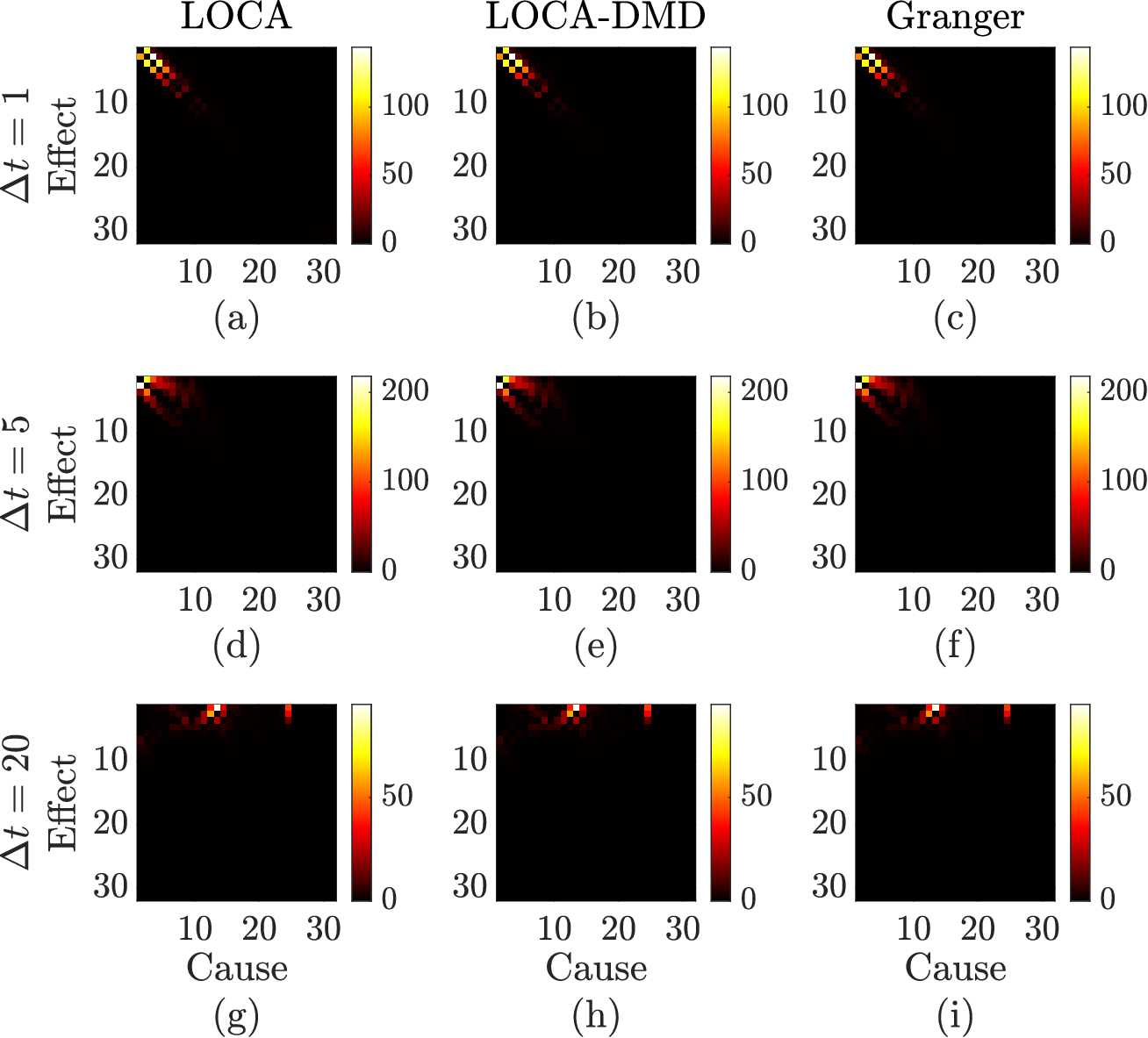}}
\caption{Causality ($F$-statistic) between POD coefficients for timesteps $\Delta t = 1$ (a-c), 5 (d-f), and 20 (g-i), computed using LOCA (a,d,g), LOCA-DMD (b,e,h) \revone{single-lag} Granger analysis (c,f,i).}
\label{fig:chan1pod}
\end{figure}

\revone{The comparatively large energy present in the $\eta$ components, shown in Fig.~\ref{fig:chanMats}(d-f), contributes directly to the LOCA statistic through the unconditional variance $\sigma_j^2$. It does not enter the standard Granger statistic in the same manner, since refitting the restricted model replaces this unconditional variance by the variance of the component of the source that cannot be linearly predicted from the remaining variables. The differences observed in Fig.~\ref{fig:chan1} are therefore consistent with the distinction established in Sect.~\ref{sec:granger}: LOCA quantifies the response to an independently specified variation, whereas Granger causality quantifies unique predictive information after conditioning on the other variables. To examine this distinction, we next transform the system to coordinates that are uncorrelated in time.} 
To show this directly, we transform our variables to uncorrelated coordinates in time. To achieve this, we use a set of proper orthogonal decomposition (POD) modes identified from running the forced linear system with a timestep $\Delta t = 1$. To compute POD, we use an inner product corresponding to kinetic energy \cite{gustavsson1986excitation}, given by
\begin{equation}
\label{eq:IP}
\langle(\hat v_1,\hat\eta_1),(\hat v_2,\hat\eta_2)\rangle = \frac{1}{k_x^2+k_z^2} \int_y \left(- \bar {\hat v}_1\Delta \hat v_2 + \bar{\hat\eta}_1\hat\eta_2\right)\mathrm{d}y.
\end{equation}
 Note that it would also be possible here and throughout this section to transform the state to velocity variables and use a standard inner product (as is done, for example, in Refs.~\cite{jovanovic2005componentwise,moarref2013model}); we have chosen to keep the velocity-vorticity coordinates so that the findings can be directly related to the constituent blocks of Eq.~\eqref{eq:chan}. 
By definition, the spatial POD modes and their temporal coefficients are orthogonal (in space and time, respectively), meaning that the time-varying coefficients corresponding to distinct POD modes will be uncorrelated.
In Fig.~\ref{fig:chan1pod}, we show the computed causality between these POD coefficients for the Couette flow system (i.e., the equivalent for Fig.~\ref{fig:chan1} but with a change of basis to POD modes). Unlike the results in Fig.~\ref{fig:chan1}, we see a close agreement between all methods in Fig.~\ref{fig:chan1pod}. 
\revone{This agreement is consistent with the analysis in Sect.~\ref{sec:granger}. For the mutually uncorrelated POD coefficients, the conditional and unconditional source variances are equal, so the LOCA and single-lag Granger measures coincide provided that the corresponding residual assumptions also hold.} 

 We can also observe that the largest causal mechanisms are highly sensitive to the timestep chosen. The leading causal interactions are confined to high energy and energetically-adjacent modes for small timesteps. At the longer timestep $\Delta t = 20$ however, the most significant causality is identified as the effect that relatively low-energy modes (particularly the 14th) have on the leading modes.

\begin{figure}
 \centering {
\includegraphics[width= 1.0\textwidth]{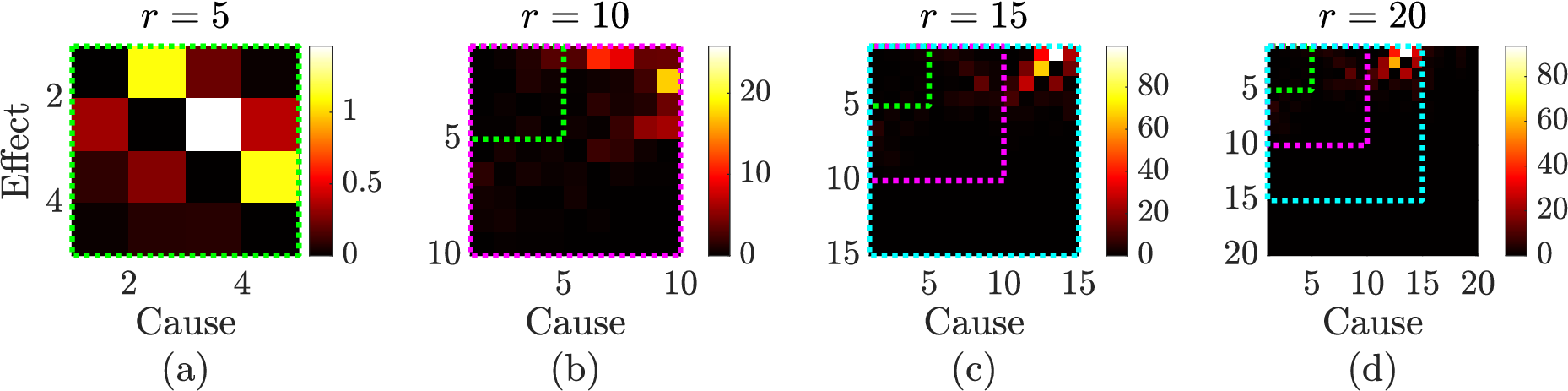}
}
\caption{Causality ($F$-statistic) between POD coefficients computed using LOCA for a timestep $\Delta t = 20$, where the dynamics are projected onto the first $r = $ (a) 5, (b) 10, (c) 15 and (d) 20 POD modes. Dotted lines indicate the truncation levels used in other subplots, highlighting the causal connections that are truncated in each case.}
\label{fig:chanPODr}
\end{figure}

 To further emphasize this point, Fig.~\ref{fig:chanPODr} shows causality results for the Couette flow system projected onto truncated sets of POD modes. When fewer than 14 modes are retained, the leading causal mechanism is not captured, potentially giving misleading findings regarding the most pertinent causal mechanisms. Note that the analysis is performed here by applying LOCA on the reduced-order linear operator. 
 \revone{LOCA-DMD applied to sufficiently rich data generated by this reduced-order operator would be expected to recover the same operator-based LOCA result. Single-lag Granger analysis need not yield identical amplitudes unless the conditional and unconditional source variances coincide and the residual assumptions discussed in Sect.~\ref{sec:granger} hold. Moreover, neither data-driven result would necessarily be the same as that obtained by applying the corresponding method to the leading coefficients from a simulation of the full-order system, since the neglected state variables would then contribute to the residual of the fitted reduced dynamics.}
 
\begin{figure}
 \centering {
\includegraphics[width= 0.65\textwidth]{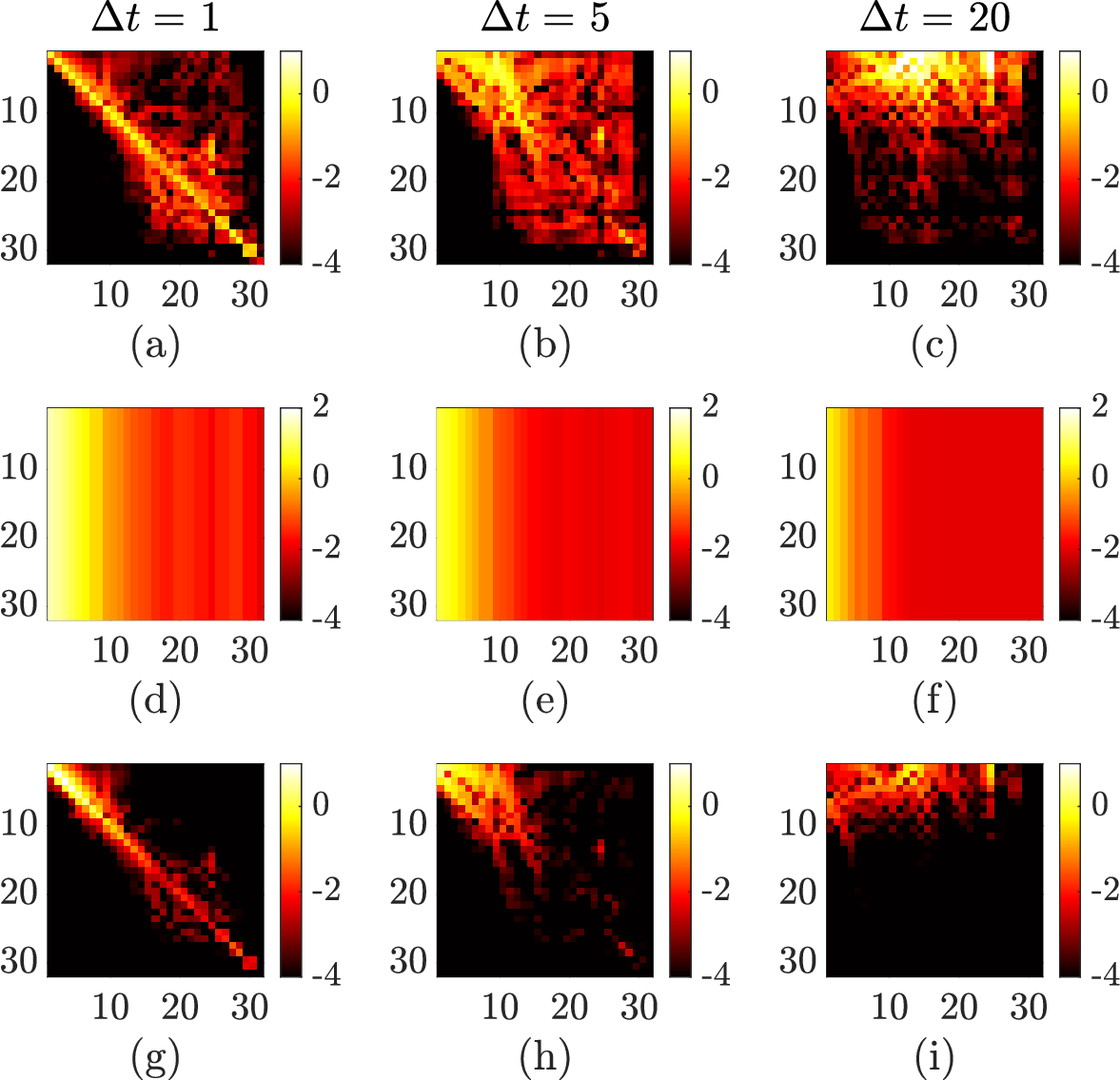}
}
\caption{As for Fig.~\ref{fig:chanMats}, but using POD coefficients: (a-c) Log amplitude of entries of discrete time operators $\log_{10}\left(\left[\tilde{\mathbf{A}}_d\right]_{ij}^2\right)$, (d-f) log of state variance $\log_{10}(\sigma_j^2)$, and (g-i) log of the product of these terms 
$\log_{10}\left(\left[\tilde{\mathbf{A}}_d\right]_{ij}^2\sigma_j^2\right)$, which is proportional to $F$ assuming uniform noise in each state.
Quantities are computed for time horizons $\Delta t = 1$ (a-c), 5 (d-f), and 20 (g-i).
}
\label{fig:chanMatsPOD}
\end{figure}

To understand this behavior in more detail, in Fig.~\ref{fig:chanMatsPOD}, we show the contributing terms to the LOCA analysis for the POD-transformed system (which is the equivalent of what was shown in Fig.~\ref{fig:chanMats} for the original variables). We observe that while the energy content of the variables ($\sigma_j^2$, shown in subplots (d-f)) decreases as expected as the POD mode index increases, this dropoff is not as large as the difference between the squares of large and small entries of the operator (subplots (a-c)). This means that the latter largely determines the size of the corresponding entries of $F$ (subplots (g-i)). In particular, this explains how we have extensive causal connections from lower energy modes to leading modes for $\delta t = 20$, shown in Fig.~\ref{fig:chanMatsPOD}(i).

\begin{figure}
 \centering {
\includegraphics[width= 0.7\textwidth]{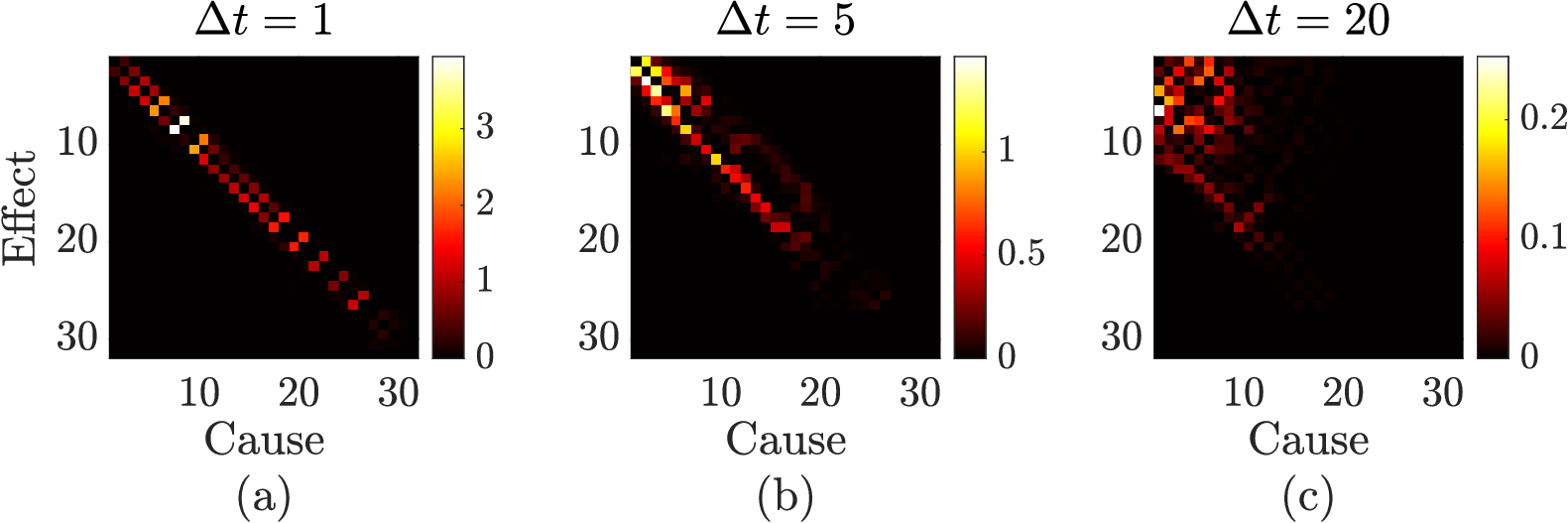}
}
\caption{Causality ($F$-statistic) identified between balancing coordinate states using LOCA for timesteps $\Delta t = $ (a) 1, (b) 5, and (c) 20. Other methods are not shown but yield similar results.}
\label{fig:chan1bt}
\end{figure}

\begin{figure}
 \centering {
\includegraphics[width= 1.0\textwidth]{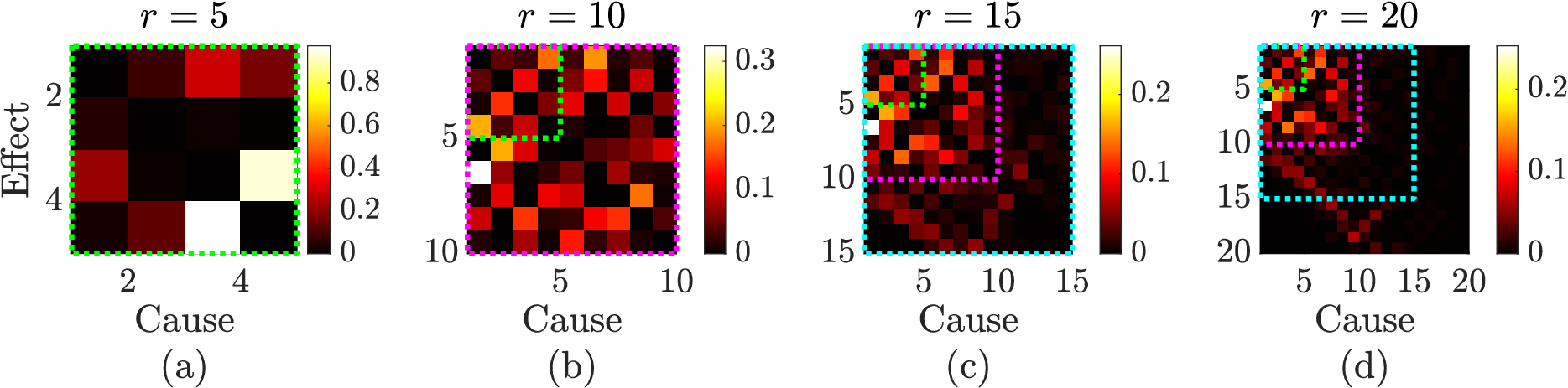}
}
\caption{Causality ($F$-statistic) identified between balancing coordinate states using LOCA, where the dynamics are projected onto the first $r = $ (a) 5, (b) 10, (c) 15 and (d) 20 state variables.}
\label{fig:chanbtr}
\end{figure}

These results suggest that the subspace identified from a truncated set of POD modes can be suboptimal for retaining causal information. This is consistent with insight obtained from other analyses of parallel shear flows. For example, it is known that it can be challenging to obtain accurate reduced-order models for linear channel flow systems using POD modes since low energy modes can still be dynamically important \citep{ilak:2008,rowley2017arfm} due to the non-normality of the dynamics. These works show that, in contrast, obtaining such model reduction via balanced truncation (or equivalently, balanced POD \citep{rowley:05pod}) can yield accurate models. 
This, along with the connection to controllability and observability discussed in Sect.~\ref{sec:bt}, motivates instead identifying such a subspace using balanced truncation of the linear system, which is designed to optimally retain the controllability and observability properties of the original system. Figures \ref{fig:chan1bt} and \ref{fig:chanbtr} show the equivalent of Figs.~\ref{fig:chan1pod} and \ref{fig:chanPODr}, but where the reduced-order dynamics are obtained using balanced truncation rather than POD. We observe that the dominant causal interactions are more heavily concentrated in the upper left blocks in Fig.~\ref{fig:chan1bt} across all time horizons, which means that the truncation of modes for the $\Delta t = 20$ case shown in Fig.~\ref{fig:chanbtr} preserves more of the causal interactions than the POD equivalent shown in Fig.~\ref{fig:chanPODr}. In particular, a truncation to 10 states retains all of the dominant causal interactions, unlike the POD case. Balanced truncation is performed on the continuous time system, so the transformation is independent of the time horizon used.

\begin{figure}
 \centering {
\includegraphics[width= 0.5\textwidth]{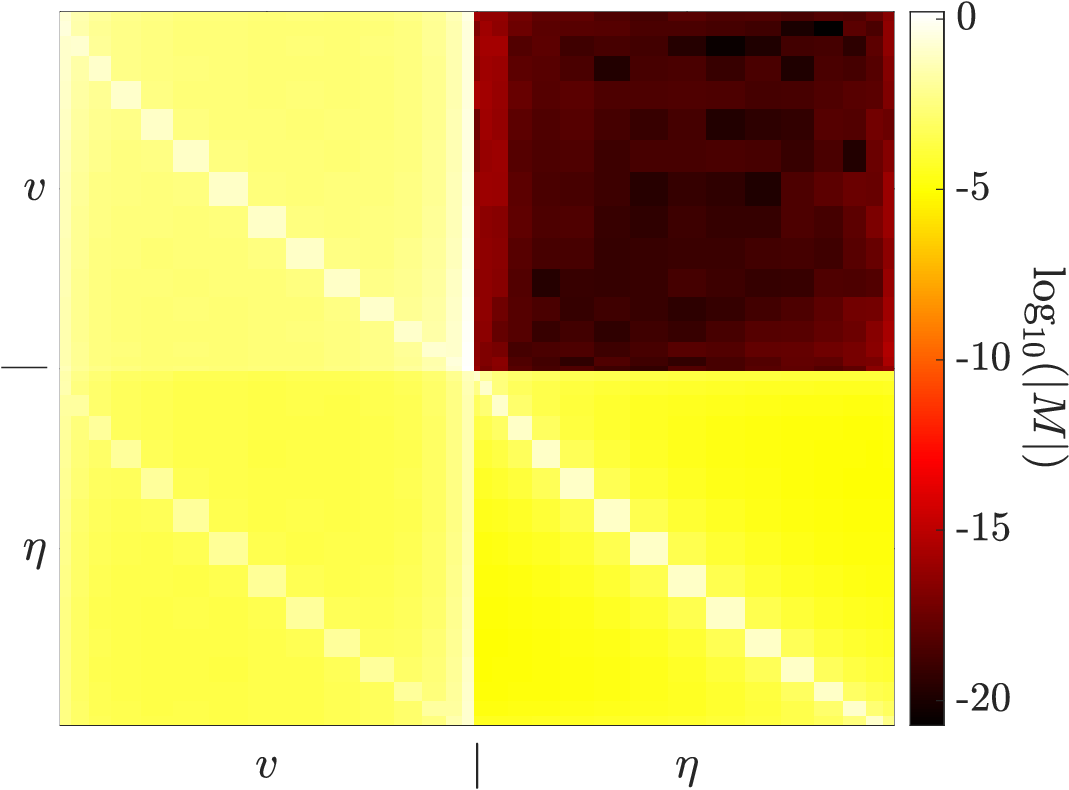}
}
\caption{Logarithm of the magnitude of the global causality indicator $\mM_\alpha = (\alpha\mI -|\mA|)^{-1}$ for the linear Couette flow system, 
\revone{with $\alpha=\gamma\rho(|\mA|)$ with $\gamma=1.05$.}}
\label{fig:chanM}
\end{figure}

Thus far, this example has focused on quantitative causality assessments over a specified time horizon. We now consider the global causality metric $\mM_\alpha$ defined in Eq.~\eqref{eq:M}. In Fig.~\ref{fig:chanM}, we compute this quantity in the original ($v,\eta$) coordinates of the system. The top right block is zero (to numerical precision), indicating no causal connection from $\eta$ to $v$. This is expected since the lower block triangular structure of $\mL$ in Eq.~\eqref{eq:chan} is retained in the discretized system Eq.~\eqref{eq:chanDisc} for all time horizons. Conversely, $\mM_\alpha$ identifies causal connections between all other state variables.

\section{Application: Two-dimensional wake flow}
\label{sec:results-wake}
\begin{figure} [htp]
    \centering
\includegraphics[trim={0cm 0cm 0cm 0cm},clip,width=0.8\textwidth]{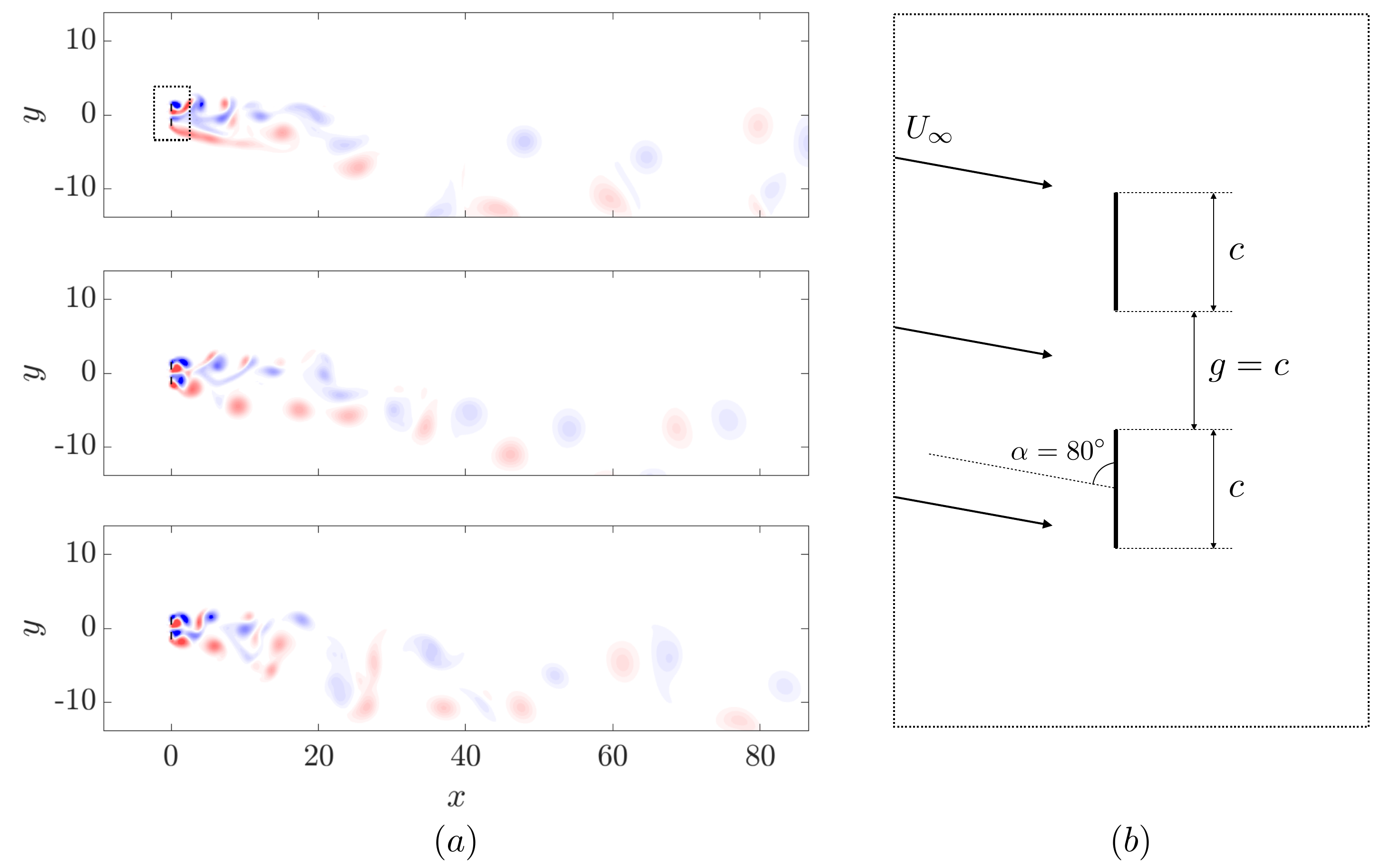}
    \caption{(a) Snapshots of the instantaneous vorticity field for the two plate flow, and (b) zoomed-in diagram of the geometry in the region identified by the dotted box in the top subplot of (a). }\label{fig:diag}
\end{figure}

We now apply the proposed methodology to analyze causality in a more complex flow featuring nonlinear dynamics. 
In particular, we analyze a flow over two collinear plates of chordlength $c$, inclined at an angle of $\alpha = 80^\circ$ to the freestream, as shown in Fig.~\ref{fig:diag}. The gap $g$ between the plates 
is fixed at one plate width, i.e.,~$g=c$. The Reynolds number based on the freestream velocity ($U_\infty$) and a single plate width ($c$) is 100.
Direct numerical simulations are performed using an immersed boundary method~\citep{taira2007immersed,colonius2008fast}.

The computational domain has an extent of $96c \times 28c$, with the inlet boundary placed $9c$ upstream of the plates. The computational domain consists of four nested grids. The finest grid, which surrounds the plates, has a spacing of $0.02c$, with each larger grid having half the resolution and double the width and height of the previous one. Representative vorticity-field snapshots are displayed in Fig.~\ref{fig:diag}, along with a diagram showing the geometrical configuration of the plates. These images show that vortices form and are shed from both the upper and lower edges of each plate. The shedding cycles are not  regular, with no persistent phase locking between the two plates. The interaction of the shed vortices produces an unsteady, chaotic wake. This configuration is identical to that considered in \cite{solera2024beta}, with Refs.~\cite{asztalos2023galerkin,solera2024beta} providing additional details concerning both the numerical setup and dynamics of such flows. Preliminary results concerning the modal decomposition and causal analysis between modes for this flow has also been considered in \cite{jimenez2026madrid}.  Although the dynamics are chaotic, the relatively small Reynolds number confines the unsteadiness to a limited range of spatial and temporal scales.

 For this example, we will focus purely on data-driven analysis (i.e.~LOCA-DMD and \revone{single-lag} Granger analyses), assuming that we only have access to the data rather than to the linearized governing equations. We use 150,000 snapshots of velocity field data, with a uniform timestep between adjacent snapshots of $\Delta t = 0.2$, non-dimensionalized by the convective timescale $c/U_\infty$.

\begin{figure}[htp]
 \centering
 \includegraphics[trim={1cm 1cm 1cm 1cm},clip,width=0.9\textwidth]{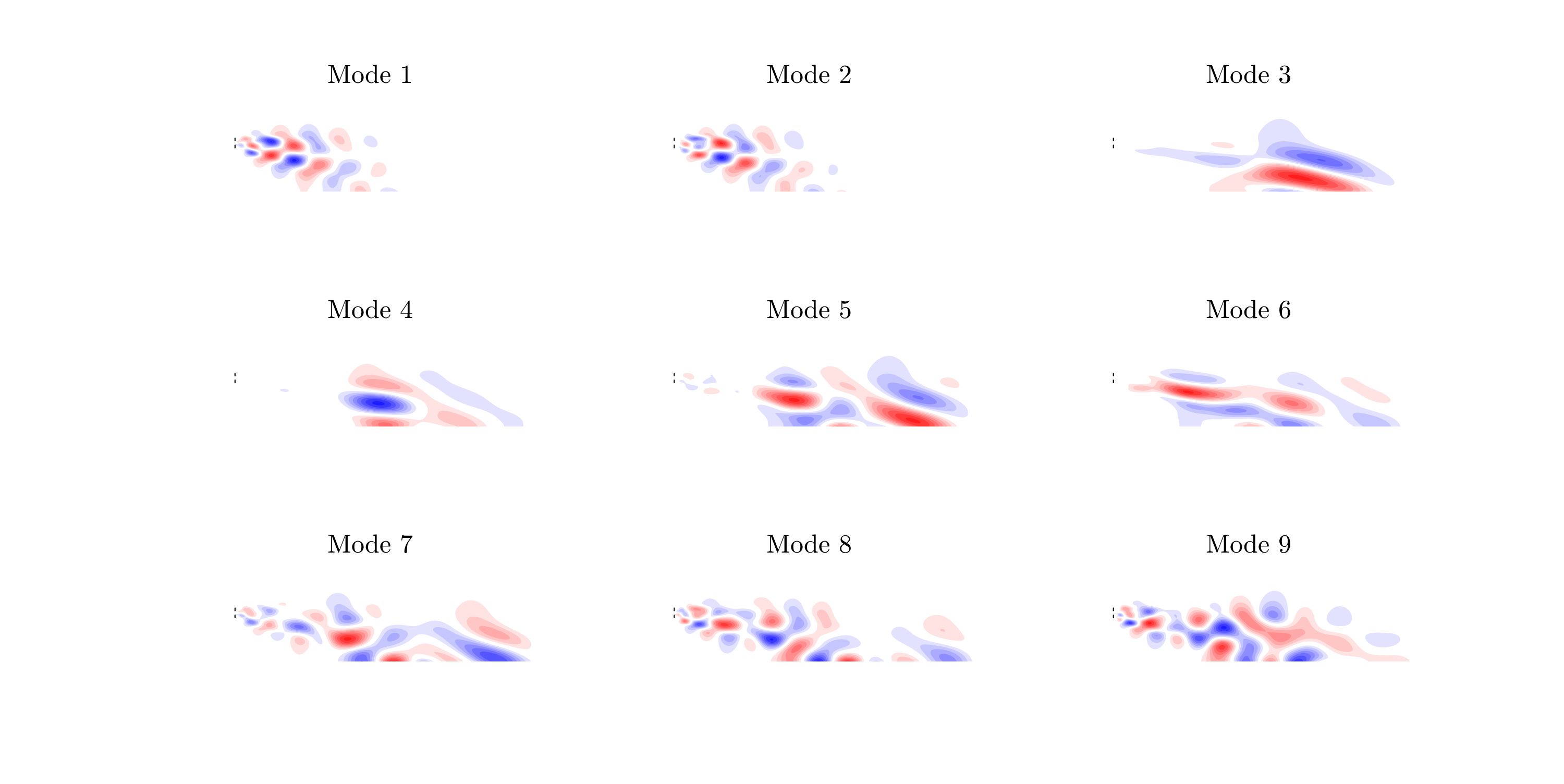}
\caption{Contours of the leading nine POD modes (streamwise velocity components)}
 \label{fig:POD_modes}
\end{figure}

The leading spatial modes obtained from performing POD on the streamwise velocity component of this dataset are shown in Fig.~\ref{fig:POD_modes}. The first two POD modes exhibit a spatial structure typical of a von K\'{a}rm\'{a}n vortex street, though their corresponding coefficients (not shown) are observed to vary in magnitude due to the chaotic nature of the wake flow. POD modes 4 to 6 represent larger structures concentrated further downstream, and are due to the meandering and intermittent behavior of the downstream wake. Modes 7 through 9 encompass a wider spectrum of spatial and temporal scales.
\begin{figure} [htp]
 \centering
\subfloat[]{\includegraphics[trim={11.5cm 0cm 12cm 0cm},clip,width=0.49\textwidth]{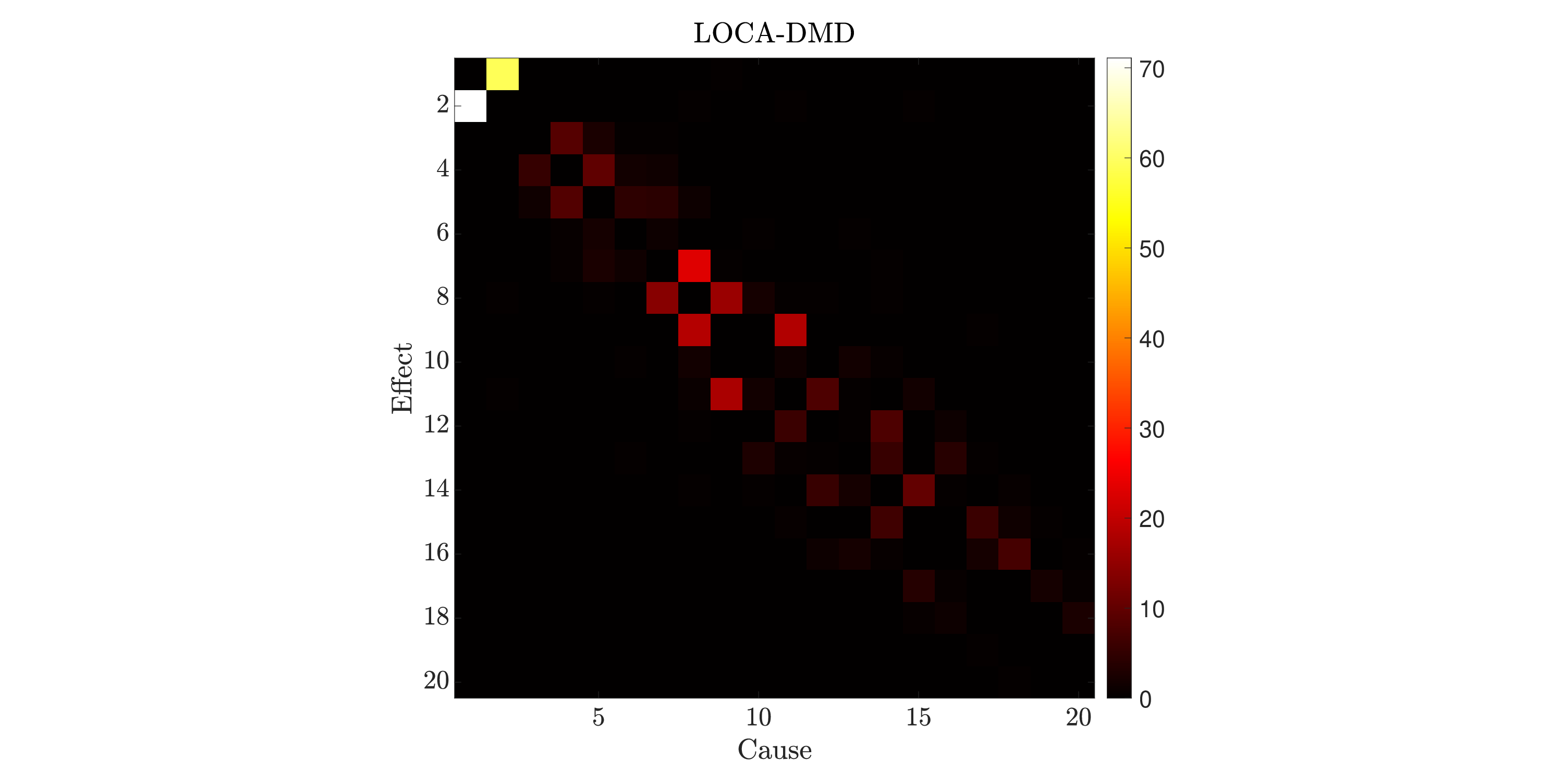}}
\subfloat[]{\includegraphics[trim={11.5cm 0cm 12cm 0cm},clip,width=0.49\textwidth]{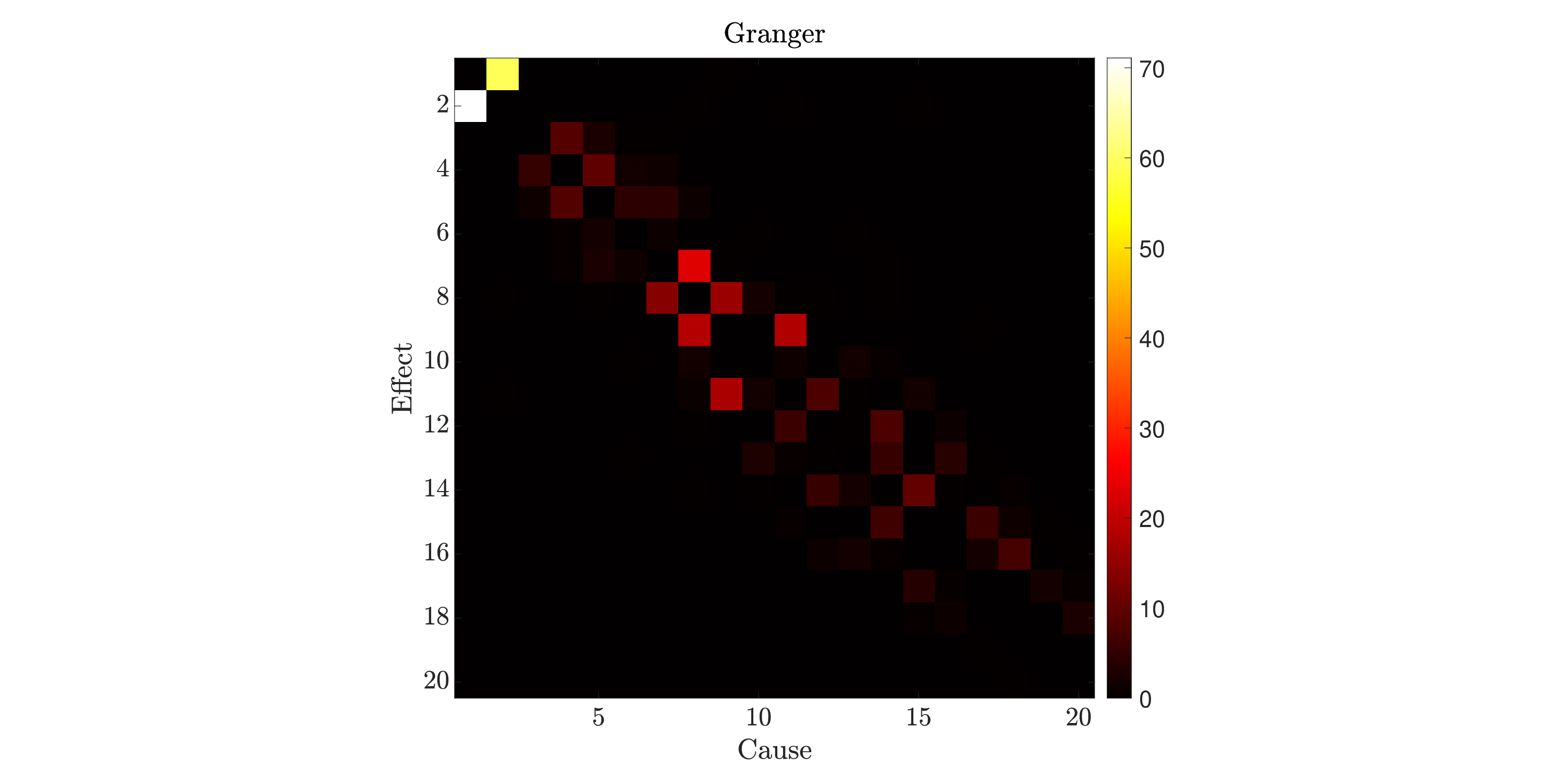}}
\caption{Pairwise causality metrics ($F$) between POD mode coefficients for $\Delta t = 1$ computed using a) LOCA-DMD and b) \revone{single-lag} Granger analysis.}
 \label{fig:POD_causality}
\end{figure}

Figure \ref{fig:POD_causality} illustrates the causal relationship between mode coefficients calculated using the Granger and LOCA-DMD methods, both computed over a non-dimensional timestep of $\Delta t=1$. As before, the diagonal self-causal elements are not plotted here or in the subsequent figures. 
A strong similarity is again evident between the results from the \revone{single-lag} Granger and LOCA-DMD analyses. Similar to the previous example, this agreement is consistent with the fact that the POD coefficients are mutually uncorrelated, so that their conditional and unconditional variances coincide. Note, however, that for this example
the error terms in the LOCA-DMD and Granger models are not due to artificially applied noise, but rather to nonlinear effects and unmodeled modes. \revone{As discussed in Sect.~\ref{sec:granger}, agreement between the two measures also requires suitable assumptions concerning the correlation between the model residual and the retained state. The present results indicate that the methods can nevertheless remain close even when this residual orthogonality is not imposed explicitly.}

In Fig.~\ref{fig:POD_causality}, we observe that the most significant causal link is between modes $1$ and $2$. This finding aligns with the fact that these modes are the most regular and alike in their spatial structure. However, this specific causal relationship provides limited novel insight, and is related to the fact that these two modes (which are approximately quarter of a period out of phase) oscillate together to represent structures convecting downstream. Other identified causal connections are predominantly situated near the main diagonals in Fig.~\ref{fig:POD_causality}, suggesting interactions occur between modes of comparable energy and  similar spatial structure. Consequently, many of these interactions are likely also a manifestation of the requirement for multiple modes to interact in a phase-locked manner in order to accurately represent traveling structures.

\begin{figure} [htp]
 \centering
 \includegraphics[trim={0cm 0cm 0cm 1.5cm},clip,width=0.9\textwidth]{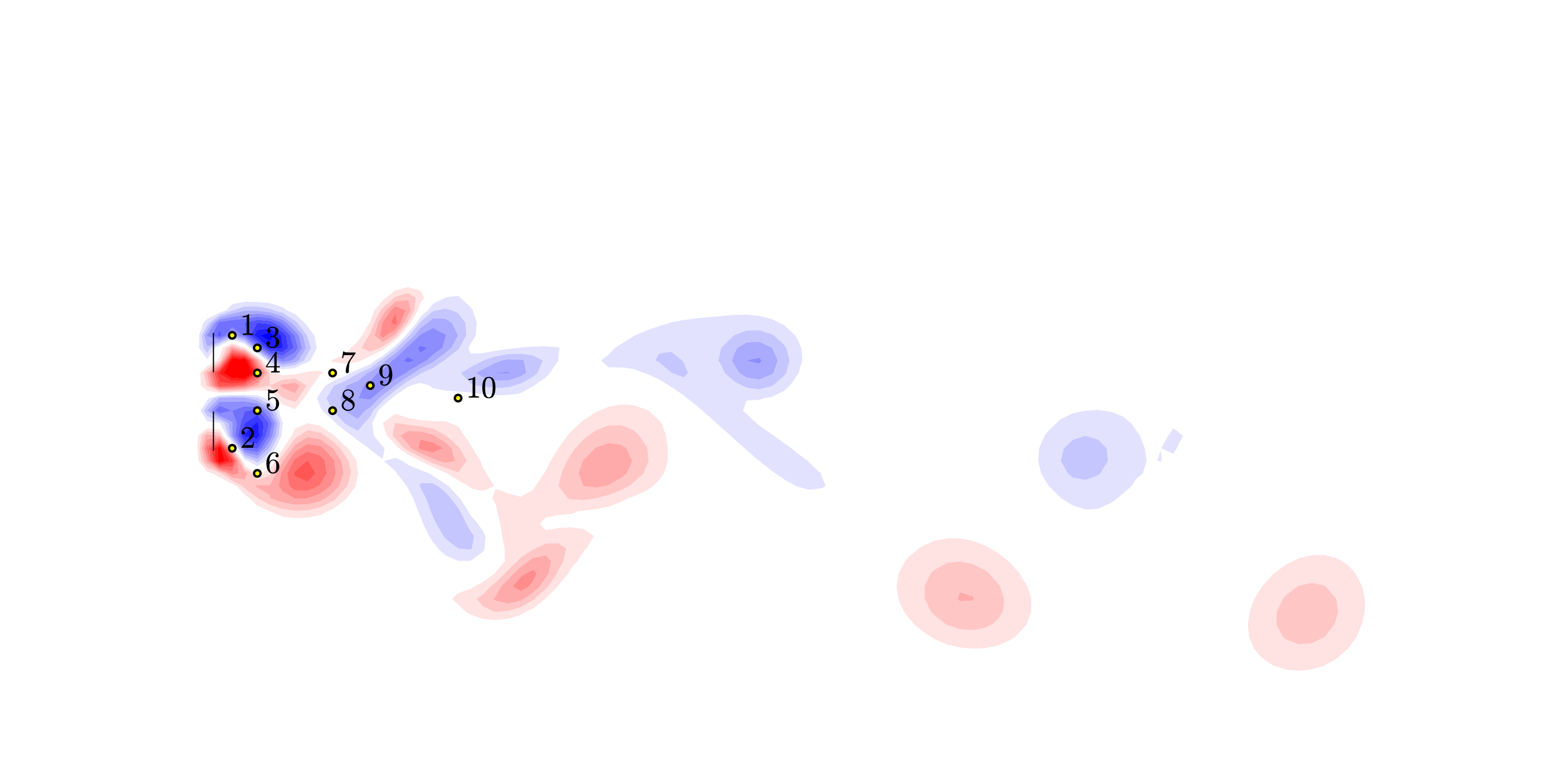}
 \caption{Locations in the domain where velocity measurements are taken.}\label{fig:Pick_points}
\end{figure}

To acquire a more insightful understanding of the flow dynamics and further test these methods, a causality analysis is next conducted directly on the original velocity variables. For simplicity in computation and visualization, a subset of $10$ discrete measurement points within the domain are selected. The locations of these chosen points are illustrated in Fig.~\ref{fig:Pick_points}. The causality assessment was performed utilizing both the streamwise ($u$) and transverse ($v$) velocity components at each of the selected points. Consequently, a total of $20$ variables were included in this analysis. 
\revone{Because these measurements are taken from a flow field that exhibits spatial coherence, substantial correlation among the retained variables is expected, making this representation a useful example of the distinction between LOCA-DMD and conditional Granger causality described in Sect.~\ref{sec:granger}.}

\begin{figure} 
 \centering{
 \subfloat[]{\includegraphics[trim={0cm 0.3cm 0cm 0cm},clip,width=0.40\textwidth]{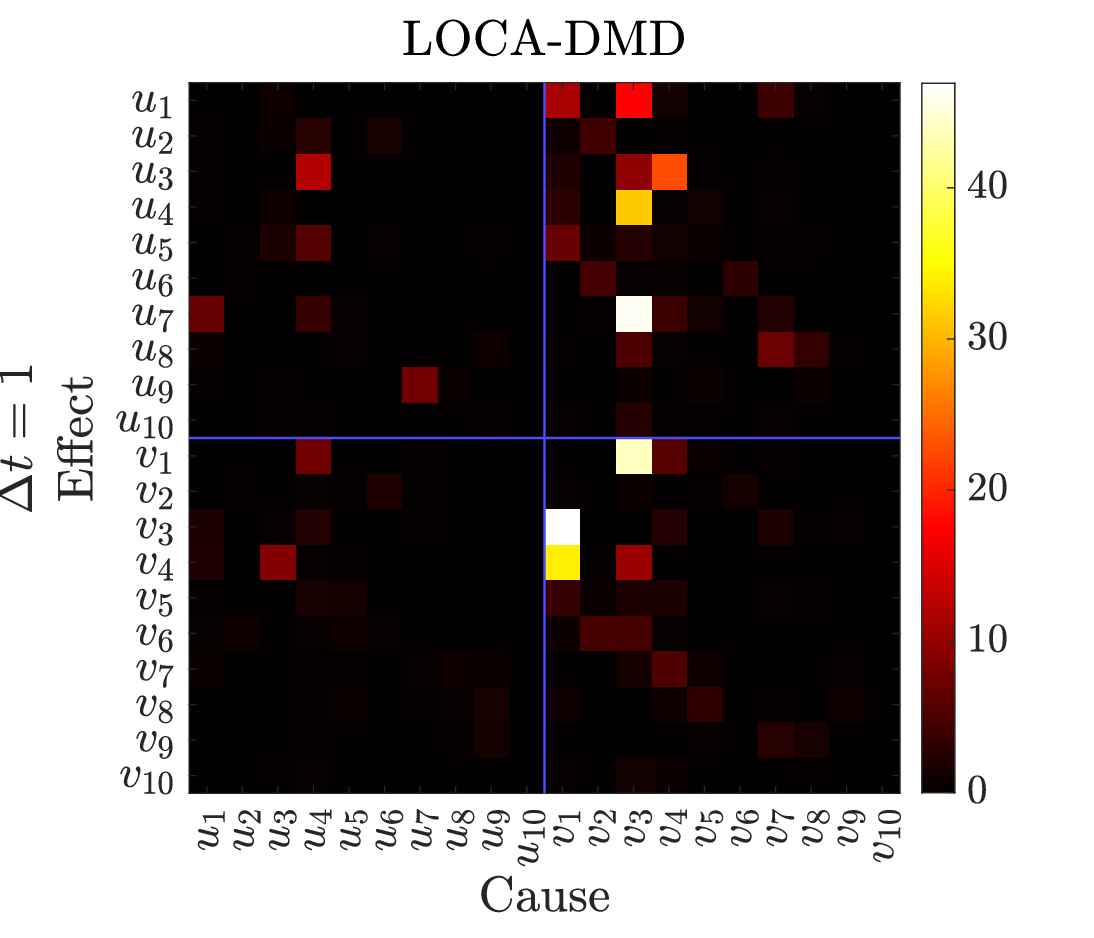}}
 \subfloat[]{\includegraphics[trim={0cm 0.3cm 0cm 0cm},clip,width=0.40\textwidth]{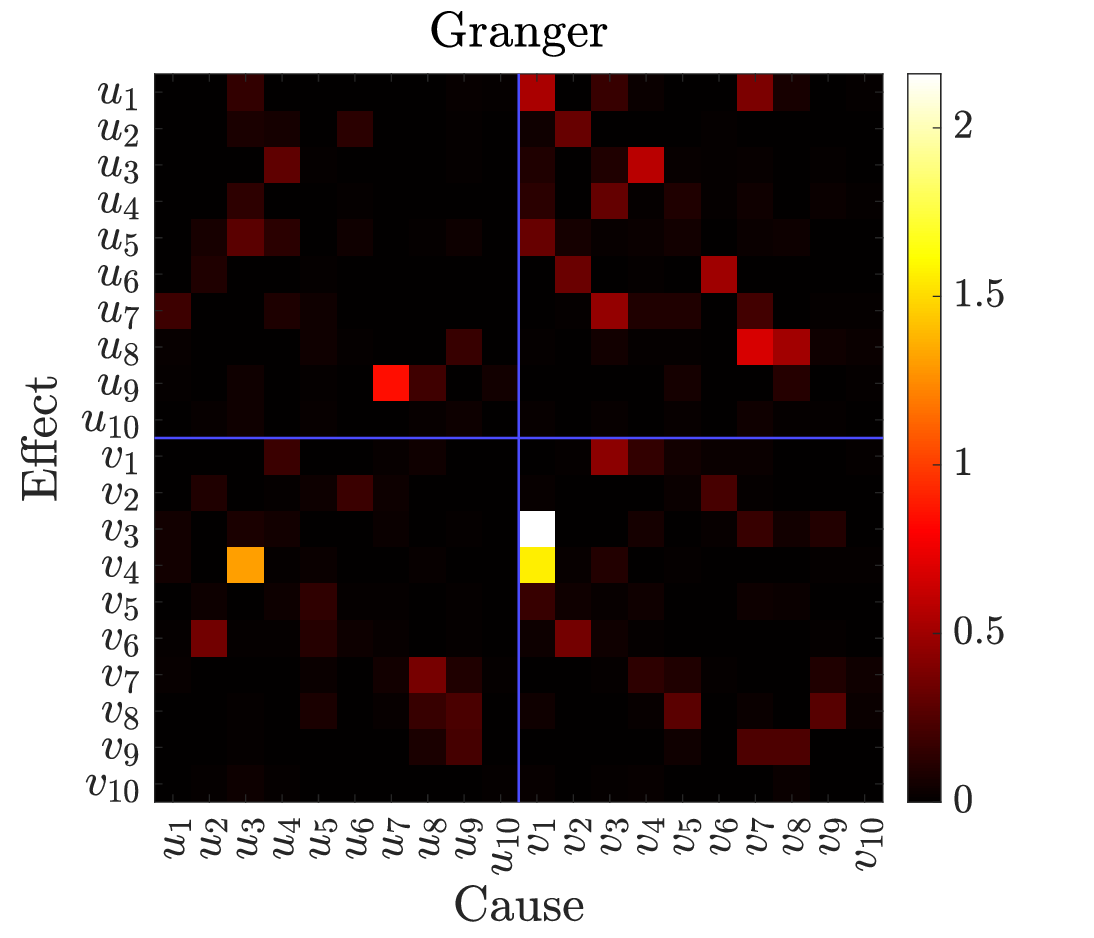}}\\
 \vspace{-0.3cm}
 \subfloat[]{\includegraphics[trim={0cm 0.3cm 0cm 1.2cm},clip,width=0.40\textwidth]{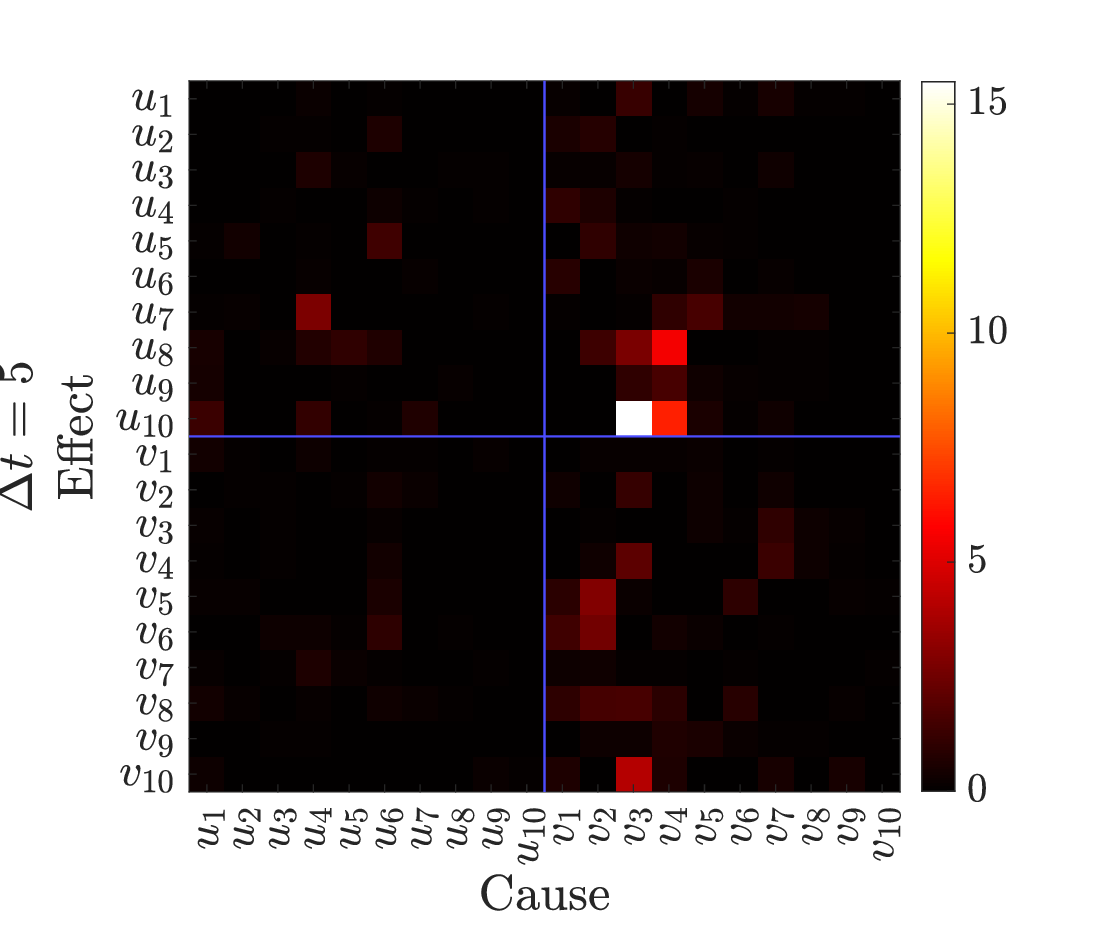}}
 \subfloat[]{\includegraphics[trim={0cm 0.3cm 0cm 1.2cm},clip,width=0.40\textwidth]{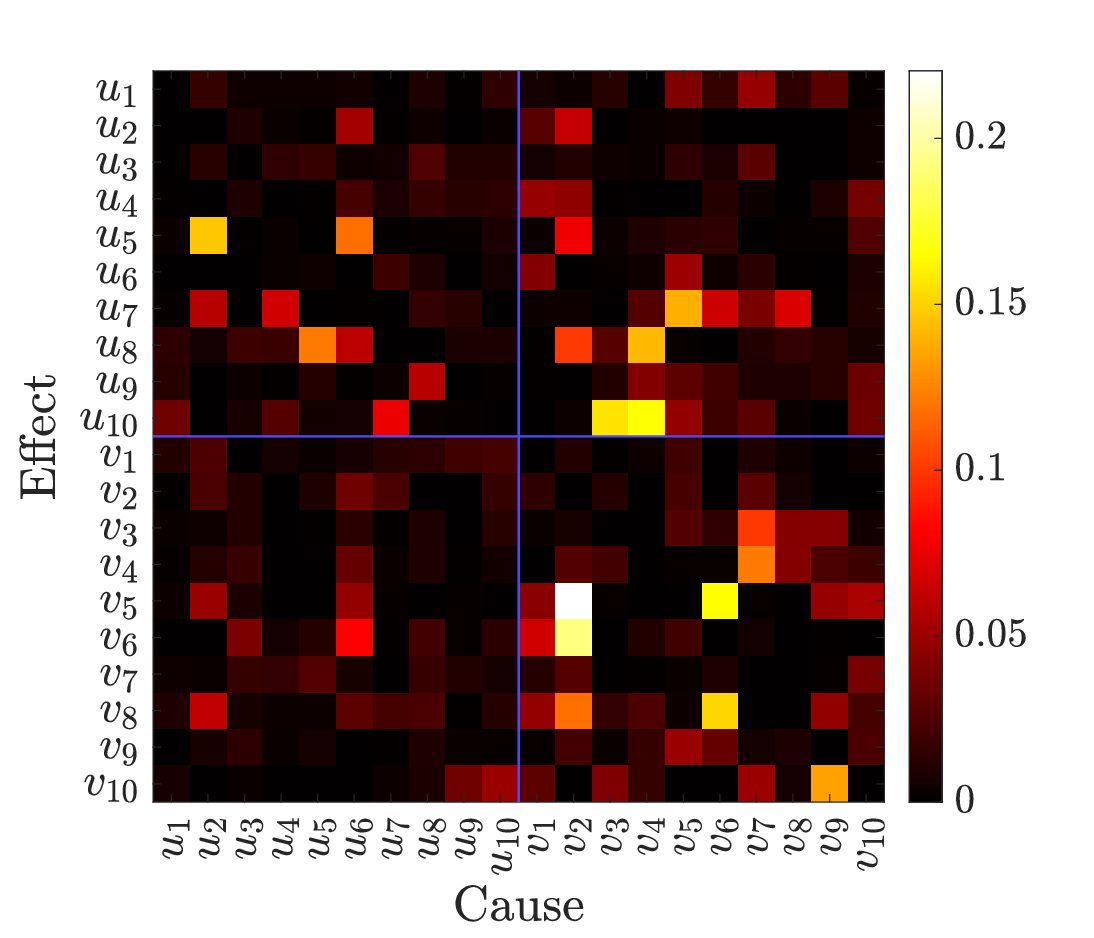}}\\
  \vspace{-0.3cm}
 \subfloat[]{\includegraphics[trim={0cm 0.3cm 0cm 1.2cm},clip,width=0.40\textwidth]{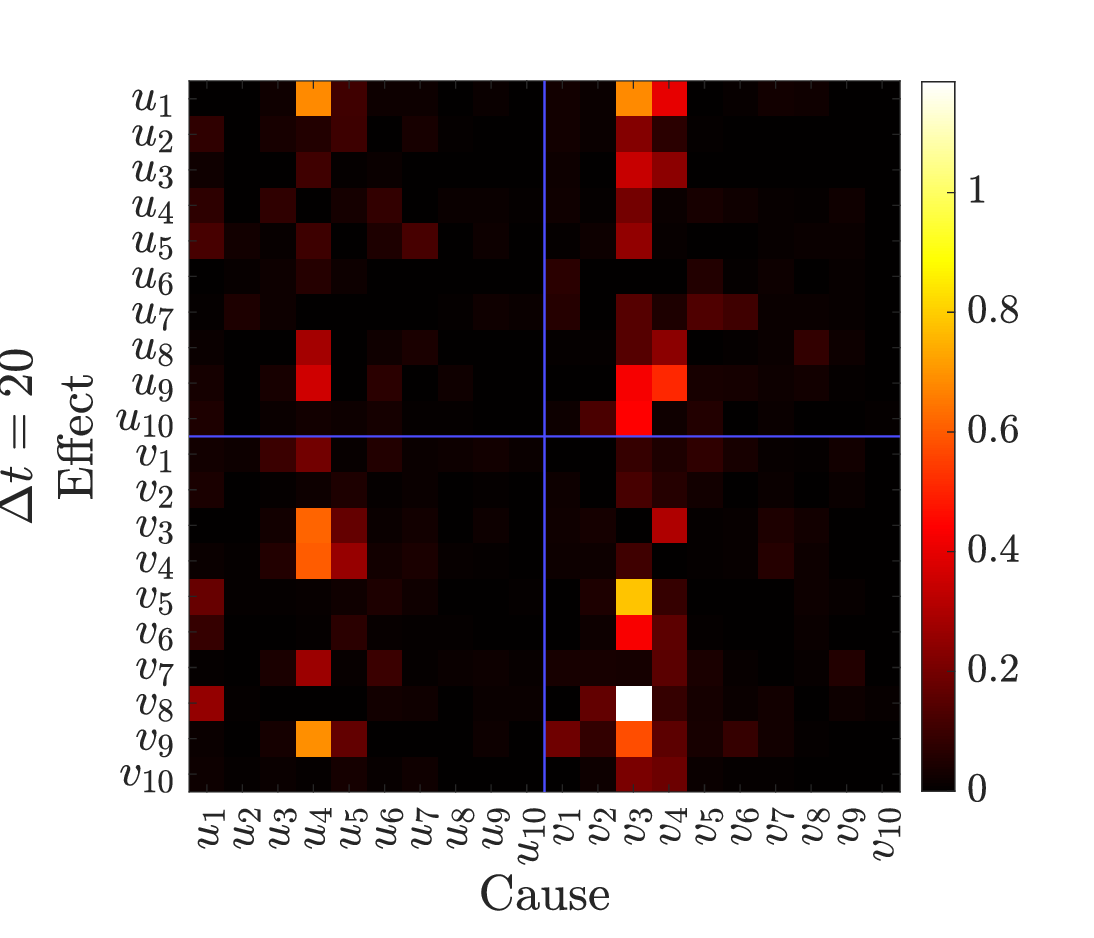}}
 \subfloat[]{\includegraphics[trim={0cm 0.3cm 0cm 1.2cm},clip,width=0.40\textwidth]{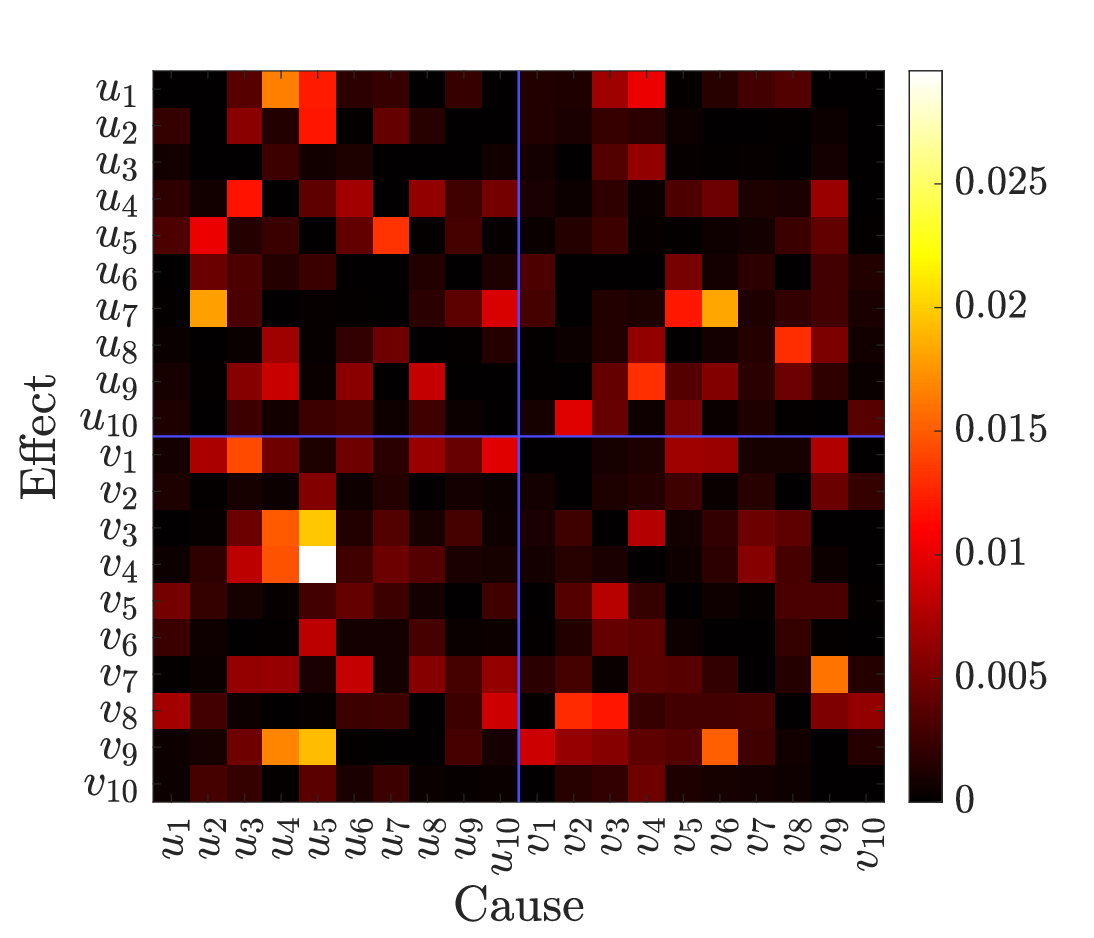}}
 }
\caption{Pairwise causality metrics ($F$) between the velocity components $u$ and $v$ at different locations, using different $\Delta t$ and comparing LOCA-DMD with \revone{single-lag} Granger Causality.}
 \label{fig:points_causality}
\end{figure}

Figure \ref{fig:points_causality} presents the results of this physical-space causality analysis, comparing LOCA-DMD and \revone{single-lag Granger causality} for various time lags ($\Delta t$). In contrast to the analysis in POD space, clear differences are observed between the two methods, as was also the case for the Couette-flow example. \revone{The velocity components at different physical locations are strongly correlated, so their conditional prediction-error variances can differ substantially from their unconditional variances. The distinction between the Granger and LOCA-DMD results is therefore consistent with the analysis in Sect.~\ref{sec:granger}; Granger causality measures the unique predictive contribution of each source after conditioning on the other variables, whereas LOCA-DMD measures the response associated with an independently specified source variation.} 
It is also notable that the magnitudes of the \revone{single-lag Granger} metrics are significantly lower than those obtained using LOCA-DMD. \revone{This is consistent with the smaller conditional source variances that can result when much of the source signal is predictable from the remaining velocity measurements. At larger time lags, the Granger values become very small and exhibit less coherent spatial structure. This may reflect both the limited unique predictive information retained after conditioning and increased sensitivity of the fitted regressions to finite-data and model-error effects. By comparison, the LOCA-DMD maps remain larger in magnitude and more spatially structured, which facilitates their interpretation as responses to independently specified variations in the measured velocity components.}

At the shortest time lag $\Delta t = 1$, the strongest causal links identified by LOCA-DMD are between points which are physically close to each other (e.g., $v_1 \to v_3$ and  $v_4$, and $v_3$ to $u_4$ and $u_7$), with the vertical velocity aft of the top edge of the top plate ($v_1$ and $v_3$) having the greatest overall causal influence on the other states. 

When the time lag increases to $\Delta t = 5$, the leading causal relationships identified by LOCA-DMD span a larger region of physical space ($v_3$ and $v_4$ to $u_{10}$). This is likely because information is able to convect further downstream over this larger time lag. For $\Delta t = 20$, the causal structure becomes remarkably distinct and clear. The primary sources of causation are now dominated by positions $3$ and $4$. This observation is consistent with the flow physics, as these points are strategically located in the near wake region immediately behind the top plate, and they are responsible for transferring information and momentum to the rest of the downstream domain over this extended period. 
\revone{This downstream causal structure is much more apparent in the LOCA-DMD results than in the single-lag Granger results. Under the latter measure, the contribution of each source is reduced by conditioning on the other correlated velocity measurements.}

\begin{figure}
 \centering {
\includegraphics[trim={0cm 0cm 0cm 0cm},clip,width= 0.5\textwidth]{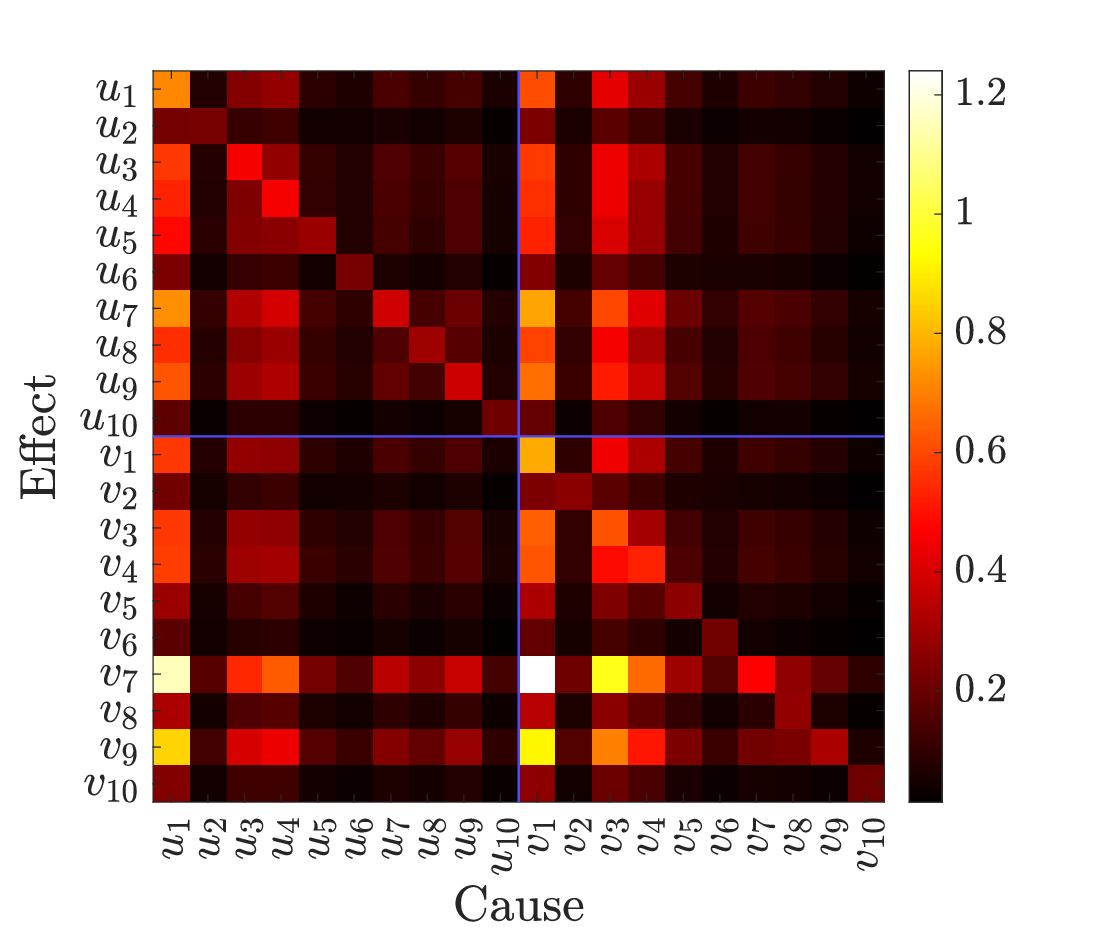}}
\caption{Magnitude of the global causality indicator $\mM_\alpha = (\alpha\mI -|\mA|)^{-1}$ for the wake flow system, with $\mA$ identified using LOCA-DMD. 
\revone{Here $\alpha=\gamma\rho(|\mA|)$ with $\gamma=1.05$.}}
\label{fig:wakeM}
\end{figure}

To estimate the causal influence over all time lags,
Fig.~\ref{fig:wakeM} shows $\mM_\alpha$ defined in Eq.~\eqref{eq:M} and computed using the $u$ and $v$ velocity components at the same locations considered previously. Over all time horizons, it is clear that that positions $2$, $6$, and $10$ exhibit negligible causal influence on other locations, which is consistent with their downstream spatial positioning (see Figure \ref{fig:Pick_points}) and the underlying flow physics. 
Conversely, positions $1$, $3$, and $4$ emerge as the most causally influential locations in the flow field. This behavior is expected given their upstream positioning near the vortex formation region, where they act as sources of information that is subsequently transported downstream.

\section{Conclusions}
\label{sec:conclusions}
In this paper, we have proposed an operator-theoretic framework for analyzing causality in linear dynamical systems, particularly those arising in fluid mechanics. These developments have been motivated in part by the utility and widespread use of physics-based linear methods in fluid mechanics \cite{schmid2007nonmodal,theofilis2011global,taira2017modal}. 
\revone{The resulting methodology, which we refer to as LOCA, provides an equation-based realization of a DCE and complements prediction-based and information-theoretic approaches such as Granger causality and transfer entropy.}

Our approach leverages the fact that for systems in the canonical form $\dot{\bc }=\mathbf{A}\bc $, the matrix exponential $\mathrm{e}^{\mathbf{A}t}$ encodes causal connections between modes at any future time $t$. We have argued that the components of this discrete-time linear propagation operator, along with knowledge of the variance of the state variables and error/noise in the discrete-time linear system, can be used to compute quantities directly related to standard causality metrics such as the $F$-statistic. 
To efficiently assess causality across all time horizons, we additionally proposed a global causality bound, characterized by the matrix $\mM_\alpha=(\alpha\mathbf{I}-|\mathbf{A}|)^{-1}$.

The operator-theoretic causality framework can be connected with several fundamental concepts in linear  systems theory. We have shown that the absence of global causality between specific modes ($M_{ij}=0$) directly implies limitations in both controllability and observability when control inputs or measurements are restricted to those modes. This relationship extends to the controllability and observability Gramians, offering potential guidance for optimal sensor and actuator placement in practical applications.

\revone{Our analysis also specifies the relationship between LOCA and single-lag conditional Granger causality. Both quantities contain the factor $\left|[\mathrm{e}^{\mathbf{A}\Delta t}]_{ij}\right|^2$, but they generally apply different statistical weightings to this dynamical coupling. LOCA uses the unconditional variance of the source variable and therefore measures the response to an independently specified variation. By contrast, a refitted Granger restricted model uses the variance of the component of the source that cannot be linearly predicted from the remaining variables, and therefore measures its unique predictive contribution. The two quantities coincide when these conditional and unconditional source variances are equal and the required residual assumptions hold. 
}

Our analysis of linearized Couette flow illustrates these distinctions. When applied to uncorrelated POD-mode coefficients, LOCA and \revone{single-lag} Granger analysis produce similar results, \revone{consistent with the equality of the conditional and unconditional source variances in these coordinates}. In correlated physical coordinates, however, the measures differ because they quantify different aspects of the source--target relationship. Notably, our analysis reveals that the most causally significant interactions can be highly sensitive to the time horizon considered and may involve lower-energy modes that significantly affect the dynamics of higher-energy structures. This finding underscores the importance of considering the full range of dynamically relevant modes when analyzing causal mechanisms, as truncated representations may omit important causal pathways. For practical applications where the system matrix is unknown, we have developed a data-driven approach based on DMD to estimate causal connections directly from time-series data. This method \revone{estimates the finite-time evolution operator and thereby extends the operator-based LOCA construction to cases in which only data are available}. The resulting LOCA-DMD method was found to agree with operator-based LOCA when directly compared in the Couette-flow example. While we formulated this data-driven variant of LOCA using DMD, it would be possible to utilize alternative methods for linear system identification, or extensions of DMD that preserve known physical structure \cite{baddoo2023physics}, are robust to sensor noise \cite{dawson2016characterizing,hemati2017biasing}, or are robust to outliers \cite{scherl2020robust,askham2022robust}. The linear Couette-flow example demonstrated that LOCA and \revone{single-lag} Granger analyses agree when the variables are uncorrelated with each other and the linear-system residual is uncorrelated with the state. The wake-flow example suggests that close agreement can also occur in practice for uncorrelated POD coefficients even when the fitted linear model represents nonlinear dynamics and its residual includes the effects of nonlinearities and unmodeled modes. \revone{For the correlated physical-space measurements, the single-lag Granger values were substantially smaller because each source was conditioned on the remaining velocity measurements. LOCA-DMD, which instead measures the response associated with an independently specified source variation, retained a more pronounced and spatially structured response over the larger time horizons considered. These results illustrate that the two methods provide complementary rather than generally equivalent descriptions of causal influence.} Overall, \revone{LOCA provides an operator-based implementation of DCE analysis for} fluid flows and other linear or linearized dynamical systems. By leveraging known or data-estimated dynamics, \revone{the method complements prediction-based causality measures and provides direct access to the finite-time response associated with specified variations of the system state}. It is hoped that insights and inferences from such analyses will assist in predicting, controlling, and designing fluid systems.

\bmhead{Funding}
AS acknowledges support from the US National Science Foundation award \#2219203. 
STMD and VJ acknowledge support from the US Department of Energy award DE-SC0025597. 
STMD acknowledges support from the US National Science Foundation award \#2238770.
LNC acknowledges support from Air Force Office of Scientific Research Grant FA9550-23-1-0269 monitored by Dr. Gregg Abate. Part of this work was performed at the 2026 Madrid Turbulence workshop, supported by the European Research Council under the Caust grant ERC-AdG-101018287.

\bmhead{Declarations}

\bmhead{Conflict of interest}
The authors report no conflict of interest.

\bmhead{Ethical approval}
Not applicable.

\bibliography{library}

\end{document}